\documentclass[11pt]{mn2e}
  \usepackage{amssymb}
  \usepackage{amsmath}
   \usepackage{graphicx}
   \usepackage{color}
\newcommand{\sect}[1]{\setcounter{equation}{0}\section{#1}}

\def\al {\alpha}

\def\AAA{{\cal A}}  
 
\def\ba{\begin{eqnarray}}
\def\bam{\begin{array}}

\def\be{\begin{equation}}

\def\bi{\bibitem}

\def\B {\overline}

\def\BBB{{\cal B}}

\def\Br{\B r}

\def\BR{\B R}

\def\Bz{\B z}

\def\cd{\!\cdot}

\def\de{\delta}

\def\DD{{\cal D}}
\def\De{\Delta}

\def\ea{\end{eqnarray}} 
\def\ee{\end{equation}}

\def\ep{\epsilon}

\def \EE{{\cal E}}

\def\fr{\frac}

 \def\ga{\gamma}

\def\ha{\frac{1}{2}~}

\def\HH{{\cal H}}

\def\inf{\infty}

\def\ka{\kappa}

\def\la {\lambda}
\def\lb{\label}

\def\LLL{\left[}

\def\nn{\nonumber}
\def\nnn{\noindent}
\def\np{\newpage}

\def\om{\omega}

\def\Om {\Omega}

\def\rh{\rho}

\def\RRR {\right]}

\def\si{\sigma}
\def\sq{\sqrt}

\def\td{\tilde}

\def\th{\theta}

\def\ti{\times}

\def\tr{\td r}

\def\ts{\textstyle}
\def\tsi{\td\si}
\def\tt{\tilde t}

\def\ve{\varepsilon} 

\def\vf{\varphi }

\def\vs{\vskip 0.5 cm}

\def\ze{\zeta}

\def\1{{\it one}}

\def\2{{\ts{\ha}\!}}

\def\3 {\ts{\frac{1}{3}\!}}
\def\4{\ts{\fr{1}{4}\!}}

\begin{document}
\title{  Thought Experiments on Gravitational Forces}
\author[D. Lynden-Bell and J. Katz]{D Lynden-Bell$^1$\thanks{email:dlb@ast.cam.ac.uk
}  \,and Joseph Katz$^{1,2}$\thanks{email: Joseph.Katz@mail.huji.ac.il} \\
 \\  {\it  $^1$Institute of Astronomy, Madingley Road, Cambridge CB3 0HA, UK}
 \\
 {\it $^2$Racah Institute of Physics, Edmond Safra Campus, 91904 Jerusalem, Israel}}

\maketitle
\begin{abstract}

 Large contributions to the near closure of the Universe and to the acceleration of its expansion are due to the gravitation of components of the stress-energy tensor other than its mass density. To familiarise astronomers with the gravitation of these components we conduct thought experiments on gravity, analogous to the real experiments  that our forebears conducted on electricity. By analogy to the forces due to electric currents  we investigate the  gravitational forces due to the flows of momentum, angular momentum, and energy along a cylinder.   Under tension the gravity of the cylinder decreases but the 'closure' of the 3-space around it increases. When the cylinder carries a torque the flow of angular momentum along it leads to a novel helical interpretation of Levi-Civita's external metric and a novel relativistic effect. Energy currents give gravomagnetic effects in which parallel currents repel and antiparallel currents attract, though such effects must be added to those of static gravity. The gravity of beams of light give striking illustrations of these effects and a re-derivation of light bending via the gravity of the light itself. Faraday's experiments lead us to discuss lines of force of both gravomagnetic and gravity fields. 
  A serious conundrum arrises if Landau and Lifshitz's definition of Gravitational force is adopted.
  
\vs\vs\vs
 \nnn PACS numbers  04.20.-q
\end{abstract} 
\np

  
\sect {Introduction}

  Magnetism is a relativistic effect of second order  in $v/c$. A  charge $q$ moving with velocity ${\bf v}$ generates a magnetic field ${\bf B}=q{\bf v\times r}/cr^3$ and the second charge $q'$ Ðmoving with velocity ${\bf v}'$ Ðis subject to a magnetic force $q'{\bf v'\times B}/c$. The ratio of the magnitude of this force to the Coulombic  electric force is of order $v v'/c^2$. How is it then that magnetism was at least as well known in ancient times as the electrical properties of rubbed amber? The total negative charge of the electrons in all known substances is so nearly balanced by the total charge of the nuclei  that they are almost electrically neutral. Indeed were this not true to an accuracy of greater than one part in $10^{36}$! then electrical interactions between the Earth, Moon and Sun would be as important as their mutual gravitation.  Thus in all bulk substances the electrical interactions are greatly suppressed by near charge neutrality,  so weak magnetic effects become comparable with the residual electrical effects due to the small charge imbalances. Magnetic effects are also enhanced by the relativistic spinning of electrons and their co-operative  alignment in the magnetic domains of suitable substances.\\
 In gravitation by contrast there are no known particles of negative mass, so for particles that move with  velocities much less than  $c$, the main gravitational interactions are the  quasi-static ones well described by Newtonian gravitation. The primary difference between electromagnetic experiments and the corresponding gravitational thought experiments stems from our ability to neutralise the Coulombic effects by using static charges of the other sign. Albeit there is evidence for an all pervasive cosmic tension in space, Einstein's $\Lambda$-term. Its gravity is currently thought to overwhelm the normal gravitational attraction at great distances and thus drive the acceleration of the universe.  The motivation for writing this article stems in part from the wish to gain a better physical understanding of how such forces work.\\
  A guiding principle of relativity  is that neither sound, nor energy, nor particles, travel faster than light in any locality. This local principle is enshrined in the dominant energy conditions (i) $T_{\mu\nu}u^{\mu}u^{\nu}\ge 0$ and  (ii) that the energy flow vector $q_{\mu}= T_{\mu\nu}u^{\mu}$ should be time-like $q^{\mu}q_{\mu}\ge 0$. Here $ u^{\mu}$ is a time-like vector corresponding to the velocity of a possible observer and we are using a metric with signature + - - - . When $T_\mu^\nu$ is diagonalisable to give an energy density $\ve$ and principal pressures $p_k$, these conditions imply $\ve\ge|p_k|$ for  $k=1,2,3$. Two oppositely directed coincident light beams give us an example of $\ve=p_1$. A straight magnetic field along $O_z$ say has a tension equal to its energy density; normal substances cannot withstand tensions approaching even $\ve/3$ but at least formally if the lambda term is treated as part of  $T_{\mu\nu}$ it has $p_1=p_2=p_3=-\ve\le0$. Here and hereafter we normally use units with $c=1$ except where we wish to emphasise the physical interpretation. 
 Einstein's equations $G^{\mu\nu}=\ka T^{\mu\nu},$ where $\ka=8\pi G/c^4$, show us that $all$ components of $ T^{\mu\nu}$ play their part in generating the metric via Newton's $G$, so they all gravitate. Newton's law of action and reaction then tells us to expect that $all$ components of a test particle's $\Delta T^{\mu\nu}=m_0u^{\mu}u^{\nu}$ should respond to gravitation via the metric through which the particle moves. In what follows we shall be particularly interested in interactions involving the spatial and mixed components of both the source of gravity and the test particle recipient.

In sections 2.1 and 2.2  we consider static gravitational forces as measured both locally and from infinity, showing that the force acts not on the rest mass but on the energy of a moving particle. The gravity of the large pressures inside spherical neutron stars can not be detected from their orbits.

In section 2.3 we consider the gravity of cylindrical shells supported against gravity by their azimuthal pressure. When such cylinders are under no longitudinal stress, increasing the rest mass increases both the gravity and the closure of the surrounding 3-space. The closure is defined as $\2[1-d({\rm circumference}/2\pi)/d ({\rm radius})] $ so the closure is zero for a plane, $ \2$ if the plane bends up to form a cylinder, and for the antipodes of the origin on a closed sphere it is one. Levi-Civita's external metric for a cylindrical space-time is
\be
d\B s^2=\B \rho^{2m}d\td t^2-\B \rho^{-2m}[\B \rho^{2m^2}(d\B \rho^2+dz^2)+\B \rho^2d\td\vf^2]
\lb{11}
\ee 
where $m$ is his mass parameter. Centrifugal force pushes a body in orbit away from the 
axis but the gravitational force on it is increased by its motion; Beyond $m=\2$ the  increased gravity due to motion overcomes the centrifugal force so all orbits hit the cylinder and circular orbits are no longer possible. Nevertheless static cylinders at $m>\2$ can still obey the energy conditions and must be considered as physical. At $m=1$ the embedding of the external space is cylindrical. Cylinders have no internal curvature
since they can be formed by rolling up a plane.  This apparent return to flat space at $m=1$ has been considered strange but, interpreted as the periodic flat space on the surface of a cylinder, it is not so bizarre. Higher values of $m$ on a cylinder without longitudinal stress would violate the energy conditions 
but, as we shall see, they become possible for cylinders under tension.

When any cylinder with a given surface density of rest mass is placed under longitudinal pressure, its gravity
increases but the closure of the space outside $decreases$. When the cylinder is under tension its gravity
decreases but its closure $increases$. This agrees with what cosmologists find when the lambda term is
considered as a somewhat bizarre form of matter with positive density but strong negative pressures in all directions. As indicated above cylinders with Levi-Civita $m>1$ can obey the energy conditions if they
are under strong longitudinal tension which of course reduces their gravity. However the closure of the external space is now increased beyond $\2$ and the cylinder's surface now has the greatest circumference of any circle about the axis. If, as in the embedding diagram  of Figure 2, we think of outwards as the direction to greater circumference then we find that the gravity of such a cylinder is outward but nevertheless toward the cylinder. Centrifugal force and gravity are helping each other in such a case so unsurprisingly there are still no circular orbits.

In section 2.4 we consider the gravity of a cylinder with a mechanical torque so that it is carrying angular momentum  up its height. The pressure tensors on the cylinder have the helical structure seen in Figure 4.
We have not found a discussion of such a case in the relativistic literature but the external metric reduces to 
Levi-Civita's when viewed in helical coordinates!  The helical structure of the external space has the 
interesting effect that without any gravomagnetic effects the orbits of zero angular momentum nevertheless
twist around as they proceed up  outside the cylinder.

Section 3 returns to the more familiar territory of the gravomagnetism due to energy currents.
After a brief, but we trust entertaining, discussion of early electromagnetic experiments, we show how
Landau and Lifshitzs' equations can be put in a form that closely resembles some of Maxwell's. 
We then treat the gravomagnetic field of a cylinder moving along itself and discuss the interesting gravity
of a beam of light described earlier by Bonnor. We treat the bending of light by working out its gravitational effect on a massive particle.

The analogy also tells us that the gravomagnetic field of a rotating cylinder is just like the magnetic field of an infinite solenoid. The return field of a long solenoid is spread over an area whose radius is comparable to the  length of the solenoid so the return field of the infinite solenoid is infinitesimal. This explains why there is no gravomagnetic field  outside an infinite   rotating cylinder and locally coordinates can be chosen with no $dt d\vf$ cross terms. Nevertheless, as Stachel (1982) pointed out , the interference of waves travelling on either side of such a rotating cylinder is affected by the gravomagnetic flux up the cylinder caused by its rotation. A similar situation occurs for the truly confined gravomagnetic field of a rolling torus. (Lynden-Bell \& Katz 2012)

Section 4 applies Faraday's picture of lines of force to the gravomagnetic flux and the gravity field and 
combines those ideas with Komar's to show their connection to Tolman's formula.

Finally in Section 5 we demonstrate a difficulty in adopting Landau and Lifshitzs' definition of what 
constitute gravitational forces in a stationary space-time.\\

	Whereas we assume astronomers are familiar with the concept of a space-time metric there may be some to whom the term 'Killing vector' is not familiar. When a space-time has a continuous symmetry such as axial symmetry then we may consider a rotational displacement that leaves the space unchanged. The
vector field of those displacements is the Killing vector. A cylinder is unchanged by uniform displacement
parallel to its axis, so the corresponding Killing vector is unit displacement of $z$. Since it is also axially symmetrical another Killing vector is displacement around the axis. Yet another is a helical displacement which combines that with a $z$-displacement. If the cylinder is static or moving along itself and rotating
about its axis which relativists call stationary, then it looks the same at all times so the system then has a
a timelike Killing vector corresponding to a displacement of time $t$ to time $t+\de t$. Any combination of
Killing vectors with constant coefficients corresponds to another symmetry and gives another Killing vector.
Thus our cylinder still looks stationary relative to rotating axes.
 \sect{Thought Experiments with static metrics}
 \subsection{Experiments of Priestley and Coulomb} 
 We consider first a massive spherical shell of radius $a$ with a small hole in it through which passes a cotton thread as illustrated in figure 1. We put this first as it provides a useful introduction to the difference between locally measured quantities, such as time intervals or forces, and those recorded or measured at asymptotically flat infinity, i.e. removed from the gravitational influence of the shell. This experiment may be compared with Joseph Priestley's by which he discovered in 1767 that the charge on an electrically conducting spherical shell with a small hole, resided on the external surface of the shell (Whittaker 1955). Except in the immediate vicinity of the hole, Priestley showed that there was no detectable charge on the inner surface.  From this finding he correctly deduced that the force between charges must obey an inverse square law. Only for that law does Newton's theorem of no field within a uniform spherical shell hold (Newton 1686).  Priestley's finding preceded the direct experiments of  John Robison who in 1769 found the repulsive force between like charges proportional to $r^{-2.06}$ while the attractive force between unlike charges varied with a power somewhat less than 2. His conjecture that the inverse square law was probably correct for both was confirmed by C.A.Coulomb  who independently reinvented Mitchell's torsion balance and found $r^{-2}$ in 1785. \begin{figure}[htbp]
\begin{center}
\includegraphics[width=10cm]{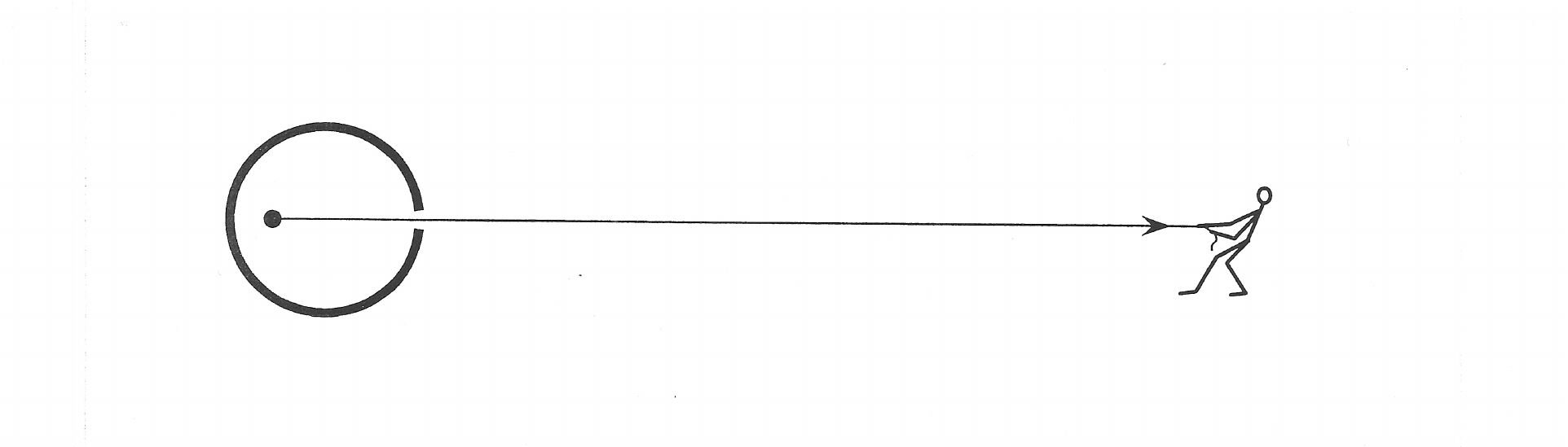}
\caption{As measured from afar a greater force is needed to accelerate a mass resting in a deep potential well.}
\label{stickman}
\end{center}
\end{figure}
  In our first gravitational thought experiment a cotton thread is attached to a small test mass $m_0$ and is gently pulled by the observer far from the shell with a tension $F_\inf$.  The acceleration  of the end of the thread is measured and compared with the tension, see Figure 1. Is $[F_\inf/$(the acceleration)] greater than, less than, or equal to $m_0$ its Newtonian value? It is not hard to give (spurious) arguments in favour of each possible outcome.\\
  1/  Mass down a potential well has lost energy and hence its mass and its inertia must have decreased; hence it should be easier to accelerate and the result shold be less than $m_0$.\\
  2/ According to Mach inertial resistance to acceleration is due to acceleration with respect to other bodies;
  this should be increased by the presence of the massive shell so the result obtained at infinity should be greater than $m_0$. \\
  3/ Space-time at infinity is flat. Space-time within the shell is flat. We should find no difference from $m_0$ the Newtonian result.\\
    In our gravitational experiment the {\color{black} space-time} metric is given by\\
 \ba
   \xi_0^2=1-2m/a;\nn;~~\xi^2=1-2m/r;~~~~~~~~~~~~~~~~~~~~~\nn\\
ds^2=\xi^2 dt^2-\ga_{kl}dx^k dx^l~~~~~~~~~~~~~~~~~~~~~~~~~~~~~~~~~~~~\nn\\
~~~~~~~= \xi_0^2 dt^2- [dr^2+r^2(d\th^2+\sin^2\th d\vf^2)]; ~~~~r\le a~~\nn\\
~~~~= \xi^2dt^2- [dr^2/\xi^2+r^2(d\th^2+\sin^2\th d\vf^2)];~~r\ge a\lb{21}
   \ea
  We shall refer to the positive definite metric $\ga_{kl}$ as the spatial metric as opposed to the metric of space-time $ds^2$. {\color{black}  Hereafter we use the unqualified term 'metric' to mean the metric of space-time and refer to $\ga_{kl}$ as the spatial metric.} The time-like Killing vector has components $\xi^\mu=(1,0,0,0).$
 Inside the sphere local time intervals are $dT=\xi_0 dt$.  So a photon emitted within the sphere with a local energy $h\nu_T$ will emerge at infinity with a red shifted energy $h\nu_t=\xi_0h\nu_T$. Evidently $\xi_0$ is the factor by which energy  down a potential well is worth less on the "International" scale. Now consider the work done by our thread when it moves  by $\de x$. In the shell this work is $F_0\de x$ in local units but in the units at infinity it is only $\xi_0 F_0\de x$; but this must be the work done at infinity $F_\inf \de x$, so the tension in the thread at infinity is less than the tension in the shell, $F_\inf=\xi_0F_0$.   We can now find the outcome of our thought experiment. Locally inside the sphere in the flat space there, force equals rate of change of momentum so $F_0=m_0d^2x/dT^2.$ Re-expressing this  in terms of $F_\inf$ and $t$ we find $F_\inf=(m_0/\xi_0)d^2 x/dt^2$ so it is harder to accelerate the string and the apparent mass as found from infinity is $m_0/\xi_0=m_0/\sqrt{1-2m/a}$.  This might be interpreted in terms of Mach's ideas on inertia induction, but contrary to that view no induced mass would be detected by an observer within the sphere, so the idea that the rest of the universe induces the observed masses cannot be found in this result. \\We now turn to the gravitational equivalent of the experiments of Robison and Coulomb; What is the force law by which a spherical shell of radius $a$ and mass $m$ attracts a test mass forced to remain at radius $r$?  There is now some ambiguity in what should be called the radius. Our $r$ is the circumference$/(2\pi) $ or [area$/(4\pi)]^\2$  but alternatives are the isotropic coordinate radius $\Br=\2[r-m+\sqrt{r(r-2m)}]$ or the radius $\td r$ as measured radially from the centre. Not surprisingly most results are much simpler when expressed in terms of $r$ rather than the others so we shall use $r$. \\
 In relativity it is common practice to eschew the concept of gravitational force fields because freely falling coordinates can be used in which such forces are absent. However at least for the special cases of static and stationary space-times such fields are well-defined and useful.  A body held static at radius $r$ has a 4-velocity $u^\mu=\xi^{\mu}/\xi$. Its 4-acceleration is non-zero despite it being  static because $du^{\mu}/d\tau =u^{\nu}D_{\nu}u^{\mu}=D^{\mu}\ln\xi$. Multiplying by $m_0$ we find the  4-force necessary to keep it static. The gravity field must be minus this $/m_0$, so, lowering the resulting gravity  3-vector with the positive definite metric gamma, we find the covariant gravity field is ${\bf\EE}=-{\bf\nabla}\ln\xi=-m{\bf \hat r}/[r^2(1-2m/r)]$ where $\bf{\hat r}$ denotes the unit vector ${\bf r}/r$. The locally measured vierbein or physical component is $(\ga_{rr})^{-\2}\EE_r=-m/[r^2 \sqrt{1-2m/r}]=-GM/[r^2\sqrt{1-2m/r}]$. The tension at infinity in a thread that holds up the static test mass is $\xi$ times the locally measured tension i.e.  $Gm_0m/r^2$. This  remains finite even when the Schwarzschild metric is due to a black hole rather than a static shell. Indeed 
 when evaluated at the hole it gives what is called the gravity of the hole. Of course this result is replaced by zero if the test mass lies inside the shell. Notice that while the tension in the thread is not constant nevertheless the red-shift-corrected tension is constant. We should remark here that what we have been calling the mass of the shell is the externally perceived mass not the sum of the mass elements on the surface but rather $4\pi a^2\sqrt{1-2m/a}(\si+P_{\vf}+P_{\th})$ where the latter two line-pressure terms are equal and $\si$ is the surface density of matter. Line-pressure gives force per unit length. This is the Tolman (1934) expression for the external mass in terms of the surface mass density $\si$ and the line-pressures, see section 5. For the static spherical shell $\ka\si= 2(1-\sqrt{1-2m/a})/a; ~~\ka P_{\vf}=  (1-\sqrt{1-2m/a})^2/(2a\sq{1-2m/a})$. The dominant energy condition is satisfied provided $a/m\ge25/12$.
 More generally if there is a spherical distribution of mass as in a spherical galaxy, the metric will be of the form $ds^2=\xi^2 dt^2-[e^{2\la(r)}dr^2+r^2(d\th^2+\sin^2\th d\vf^2)]$. All the above results hold except that  
  $\ga^{rr}=e^{-2\la} = 1-2m(r)/r$ where $m(r)=Gc^{-2}\int_0^r 4\pi r^2 \rho(r)d r$ gives the mass distribution inside the galaxy and $\xi^2$ is no longer $[1-2m(r)/r]$; rather ${\bf\EE=-\nabla}\ln\xi=-[m(r)+4\pi G r^3p(r)/c^2]{\bf\hat{r}}/[r(r-2m)]$. Here $p(r)$ is the pressure due to stellar motions here assumed to be isotropic. Notice that it is the  pressure that is acting here as a source of gravity a totally non-Newtonian effect. Also $4\pi r^3/3$ is not the volume within $r$. We have not found a precise convincing physical interpretation of this form of source term which was discussed  by Ehlers {\it et al.}(2005). The gravity of such pressure terms is sizeable inside neutron stars however despite their effect on the internal structure
the external fields of spherical stars are determined solely by the mass which can be expressed as an integral over the density alone vis.  $M=\int 4 \pi r^2\rho dr$. Thus the contribution of pressure to the gravity is not detectable in the external field nor in the orbital parameters.
  The same arguments may be applied to any static metric and give a covariant gravitational force $m_0{\bf\EE}$ where ${\bf\EE=-\nabla}\ln\xi$. However, denoting unit vectors by a hat, the magnitude $m_0\ga^{kl}{\hat \EE_k} \EE_l $ of the vierbein  vector  depends on the metric. The tension in the thread is still $\xi$ times this. The  taught  thread necessary to hold the particle statically will lie along a geodesic of the spatial metric $\ga_{kl}$. Since $m_0$ is a small test mass, we do not have to account for the disturbance to the metric due to the thread in calculating this tension.\\Returning to our spherical shell but now including Einstein's dark energy $\Lambda$ term we have $\xi_0^2= 1-2m/a-\Lambda r^2/3; ~~\xi^2=1-2m/r-\Lambda r^2/3$
and deduce the gravity field outside the shell ${\bf\EE}=(-m/r^2+2\Lambda r/3){\bf\hat r}/\xi$
This can be interpreted as the attraction of the massive shell plus a repulsion due to a uniform distribution of dark energy. Notice that the latter force is present even when there is no shell.
\subsection{Gravitational force on a moving particle}
  Radial lines are globally and locally straight so motion along them involves no acceleration due to  spatial curvature. We take a particle forced to move radially at constant speed,$v=\sqrt \ga_{rr}dr/(\xi dt)$. Its 4-velocity is $u^{\mu}=(1-v^2)^{-\2}[\xi^{\mu}/\xi+(0,v/\sqrt\ga_{rr},0,0]$. We again calculate its  4-acceleration and the 4-force necessary to keep it going at constant velocity. Again this must be equal and opposite to the force of gravity. The 4-force needed to keep it at constant speed is $F^{\mu}=d(m_0u^{\mu})/d\tau$
  while the 3-force is $f^k=\sqrt{1-v^2}F^k$, so again lowering the index on the 3-vector $f^k$ via $\ga_{kl}$
  we find $-f_k=(m_0/\sqrt{1-v^2})\EE_k$, where $\EE$ is the gravity field given above. It is the particle's energy rather than its rest-mass is pulled by gravity.\\
  Our next experiment is on a test particle forced to move at constant angular velocity in a circle in the equatorial plane. As for the static particle this motion is along a Killing vector $\ze^{\mu}$ which is a combination of $\xi^{\mu}$ and the angular Killing vector $\eta^{\mu}$. In the equatorial plane $\eta^{\mu}\eta_{\mu}=-r^2$ and as in our first experiment  $u^{\nu}D_{\nu} u^{\mu}=D^{\mu}\ln\ze$. From this we derive the 3-force necessary to keep it so moving,
\be
f_k=\sqrt{1-v^2}~m_0\partial_k\ln\ze=\frac{m_0}{\sqrt{1-v^2}}[\partial_k\ln\xi-\de_k^1v^2/r]
\lb{22}
\ee
Evidently the second term in the square bracket gives the centripetal force and the first term gives minus the gravitational force. Of course when $v$ is so chosen that gravity provides all the centripetal force necessary then the force $f_k$ that we have to impose from outside will vanish and the particle proceeds along its circular orbit without ex-cathedra forcing by us. 
For free geodesic motion in Schwarzschild space-time $(1-2m/r)\dot{ t}=E, ~ r^2\dot\vf=h,~\dot s^2=1$, so
\be
\dot r^2= E^2-(1+\fr{h^2}{r^2})(1-\fr{2m}{r}) 
\lb{23}
\ee
 thus the centrifugal potential gradient is $h^2r^{-3}(1-3m)$ which changes sign at $r=3m$. Closer in the gravity on the energy of the angular motion exceeds the normal centrifugal force and greater angular motion only increases the infall rate.  
 For a static metric with axial symmetry the azimuthal Killing vector's magnitude $[-\eta^{\mu}\eta_{\mu}]^{1/2}=R$ defines a circumferential radius which we use as our first spatial coordinate. For all such metrics our former argument still holds so the covariant 3-force necessary to keep our test particle in circular orbit at $R$ with speed $v$ is $f_k=\frac{m_0}{\sqrt{1-v^2}}[\partial_k\ln\xi-\de_k^1v^2/R]$.\\
  The gravitational force on a particle  in general motion can only be decided once the contribution of the inertial curvature terms have been removed, as seen above for circular motion. The local curvature of the path in space gives the departure from a spatial geodesic, but, as we shall see later, the spatial geodesics in a curved space, although locally as straight as possible, are not globally straight. Such a spatially geodesic path transiting a curved space region emerges to a large distance in a direction different from that at which it entered. We defer the question of how momentum changes along general paths through space-time, as this involves discussion of these deeper issues which are explored in Section 5. For the present we shall consider only paths that are either straight by symmetry or along Killing vectors for which global considerations apply.
 \subsection{The Gravity of pressure or tension}
  There are three different types of  conserved current in general relativity. These are the currents of 
 energy, momentum  and angular momentum. The energy current is antisymmetric to time reversal and can only be discussed in non-static metrics, but the other two are symmetric under time reversal. The momentum current up a static cylinder with its axis vertical is directly related to the pressure across a cut at constant $z$. This gives the flow of upward momentum upwards which is the same as the flow of downward momentum downwards. Of course if that pressure is negative corresponding to a cylinder in tension this gives  a downward flow of upward momentum etc. Likewise if  we clamp the upper end of a solid cylinder and then twist the lower end about the axis we will produce an elastic torque in the cylinder. This may be viewed as carrying positive angular momentum upwards while carrying downward angular momentum downwards. The gravitational effects of these latter two currents can  be studied via thought experiments using static pressurised or torqued cylindrical shells.
 Before we consider the gravitational effects of cylindrical wires that carry currents of mass-energy analogous to the electrical experiments of Oersted and Ampere, we shall  discuss the gravity fields of static cylinders. While the metrics of infinite cylinders can be readily derived from Einstein's equations,  nevertheless their physical interpretation in general relativity is less lucid as the metrics can no longer be flat at infinity. For this reason we shall also discuss the metrics of long thin prolate spheroids, which, near their equators, approximate those of cylinders but nevertheless become flat at infinity. 
In classical gravitation the potential of an infinite cylindrical shell of radius, $b$, and mass per unit length, $\td\mu$, is $\psi=\psi_0 - 2G\td\mu\ln (R/b); R\ge b$ and within the cylinder it is just $\psi_0$. At large $R$ this potential diverges logarithmically; the potential $\psi_0$ at the central point of a finite cylinder of length $2a$ is $2G\td\mu\ln[\sqrt{1+a^2/b^2}+a/b]$. As expected this diverges as $a\rightarrow\inf$.\\
This central potential of a finite cylinder is closely related to the constant internal potential of a prolate spheroidal shell of semi axes $\sqrt{a^2+b^2},~ b$, made by taking two concentric similar spheroids of slightly different sizes and filling the shell between them with a constant density. As Newton showed, its potential is constant internally; it is given,  internally and externally by the expressions,
 \ba
 \psi= \psi_0=\frac{GM}{a}\ln(\sqrt{1+a^2/b^2}~+a/b);~{\tilde r}\le b\nn:\\
 ~\psi= \frac{GM}{a}\ln(\sqrt{1+a^2/\tr^2}~+a/\tr);~\tr \ge b,
 \lb{24}
 \ea
where $\tr^2$ is the larger of the two roots for $\la$ of $\la^2-(R^2+z^2+a^2)\la +a^2R^2=0$. $\tr$ is constant on spheroids confocal to the shell.  $\psi$ can also be written in terms of $\Br=\sqrt{\tr^2+a^2}$ in the form $\psi=\frac{GM}{2a}\ln(\frac{\Br+a}{\Br-a})$. Notice that when $\tr>>a$ then $\Br\rightarrow \tr\rightarrow r,~~\psi\rightarrow GM/r$. The surface density of mass on the spheroid $\tr=b$ is $\si=M/[4\pi b^2 \sqrt{1+a^2R^2/b^4}]$.
 For long thin prolate spheroids the potential in the range $b\le\tr<<a$ is of cylindrical form near the equator
 with $\td\mu=M/2a$,
 \ba
 \psi=\psi_0-2GM/(2a)\ln(\tr/b);~~~~~~~~~\\\tr=R[1+z^2/(a^2+R^2)+O(z^4/a^4)];~~~\si=\td\mu/(2\pi b)\nn.
 \lb{25}
 \ea
    Potentials already known from Newtonian theory can be used as part of the relativistic metric.  In static empty axi-symmetric spaces Weyl showed that the metric can be chosen in the form $ ds^2=e^{-2\psi}dt^2-e^{2\psi}[e^{2k}(d\BR^2+d\Bz^2)+\BR^2d\B\vf^2]$ and with that form the time-time component of Einstein's equations becomes the flat-space $\nabla^2\psi(\BR,\Bz)=0$. Thus both outside the cylinder and the spheroid we may use the expressions from Newtonian theory given above but with $R,z$ replaced by $\BR,\Bz$. Notice however that the Weyl coordinate $\BR$ is no longer  the magnitude of the azimuthal  Killing vector but now $ \sq{- \eta^{\mu}\eta_{\mu}}= e^\psi\BR$. It is also true that the mass per unit length $\mu$ occurring in the relativistic metric is no longer the sum of the elements of rest mass but is  found as an integral over the elements of the stress tensor via Tolman's formula.  We discuss that in connection with Faraday's electromagnetic  experiments as it involves gravitational fluxes and holds for general stationary metrics. To find the remaining metric function $k$, Weyl set $D_*=\partial_{\BR}-i\partial_{\Bz}$ and gave the equation $D_*k D_*\ln \BR=(D_*\psi)^2$  which contains all the non-linearities of Einstein's theory but  is readily solved by a line integral once $\psi$ is known. For the prolate spheroid this yields (Katz {\it et al} 2011)
     $k=-\2G^2M^2a^{-2}\ln[1+(a\BR/\tr^2 )^2]$. For the cylinder we integrate Weyl's equation to find $k=4\mu^2\ln(\BR/b)+const$. For cylinders the external metric nowhere touches the axis where $k$ has to be zero to avoid a line singularity. As a result the zero point of $k$ can only be determined from the boundary conditions at the cylindrical shell. These involve not just the gravitating mass per unit length $\mu$ but also the principal surface stresses expressed as line-pressures $P_\vf$ circumferentially and $P_z$ vertically i.e. parallel to the cylinder's axis. The standard Levi-Civita external metric for a static cylinder   in dimensionless form is (\ref{11}), $d\B s^2=\B \rho^{2m}d\td t^2-\B \rho^{-2m}[\B \rho^{2m^2}(d\B \rho^2+dz^2)+\B \rho^2d\td\vf^2]$, but in this form $\td\vf$ does not in general run from $0$ to $2\pi$. We therefore take a new variable $\B\vf=C\td\vf$ with $C>0$ and that range. We also convert to dimensional coordinates by writing $\B \rho=\B R/b,~  z=\Bz/b,  bC\td t=\B t , ~ bCd\B s=ds$ where $b$ is the radius of our cylinder. This yields the metric in Weyl's form with $0\le\B\vf<2\pi$, $\B\rh\ge1$,
\be
ds^2=\B \rho^{2m}d\B t^2-\B \rho^{~-2m}[C^2\B \rho^{~2m^2}(d\B R^2+d\B z^2)+\B R^2d\B\vf^2].
\lb{26}
 \ee
 
 On a cylindrical shell at $\BR=b$ this $\Bz$ coordinate is not continuous with height $\td z$ along the axis, $\td z=C\Bz$.
 The rate of increase in circumference$/(2\pi)$ with  radial distance is then $(1-m)/C$ at $\BR=b$.  C has been called the  conicity parameter.  $\psi=-m\ln(\BR/b)$ so $m/2=\mu$, but this is amended in (\ref{212}).
 Levi-Civita's mass parameter $m$ has been called the mass per unit length but beware! even in the simplest case with no vertical pressure up the cylinder and an almost classical shell, $G$ times  the mass per unit length is not $m$ but is $m/2$. Also Levi-Civita's dimensionless $m$ does not have the same meaning as Schwarzschild's $m$, nor does it have the  dimension of a length. Despite these drawbacks Levi-Civita's
$m$ is so well established in the literature that those who write without it are likely to be misinterpreted or ignored so we shall continue to use it with the understanding that its use is restricted to cylinders. 

   To see the effect of longitudinal pressure on the gravity field of a shell we must compare experiments on metrics with the same mass distribution but different pressures. In doing this we should compare their forces at the same radii. By this we can either mean the same distance from the axis of the cylinder or on a circle with the same circumference. We choose the latter as this is the magnitude of the azimuthal Killing vector, which also appears in the angular momentum of a particle in orbit, but it is worth remarking that the detailed result does depend on what we keep fixed during the comparison. We therefore rewrite the metric in terms of the circumferential radius $R=\BR(\BR/b)^{-m}$ 
 and  $\td z=C\Bz$ which is continuous with the flat space coordinate that measures height along the axis. 
 We are unable to use $R$ as a coordinate when $m=1$ because all the external space has the same circumference as is clearly seen in Figure 2; in that special case we shall continue to use $\BR $ as our coordinate.  Setting $n=1/(1-m);~~\rho=R/b$,
 \ba
 && ds^2=\rho^{2nm}d\B t^2-[n^2C^2\rho^{2nm^2}dR^2+ R^2d\B\vf^2+\rho^{-2m}d\td z^2];\nn\\
 \lb{27}
 &&~~\psi=-\frac{m}{1-m}\ln(R/b);~~R\ge b\\
 \lb{28}
&&  ds^2=dt^2-(dR^2+ R^2d\vf^2+dz^2);~~R\le b
\lb{29}
 \ea
  Fitting the metrics (\ref{27}), (\ref{29}) across the shell using Israel's formalism adapted to stationary space-times as in  Lynden-Bell \& Katz (2012) we find, writing $\Pi_z$ for the total stress along $z$
 \ba
\td\si=\ka b\sigma=[1-(1-m)^2/C]\le1~\nn,\\~\ka\Pi_z/(2\pi)=\ka bP_z=[C^{-1}-1]~,~\ka P_\vf=m^2/(bC).
\lb{210}
 \ea
 The rest-mass per unit height measured along the axis is $2\pi b\si=4\td\si/G$. The Komar mass per unit height is $\2 m/C$. All these agree with the results of Bi\v{c}\'{a}k \& \v{Z}ofka (2002)   when our normalisation of $R$ is accounted for.
 For $m>1$ and $\BR>b$ "outside" the shell, we see that the circumferential radius $R$ DECREASES as 
 $\BR $ increases so $R$ attains its greatest value $b$ on the shell. This is reminiscent of spaces nearing closure and is more easily understood with the aid of the embedding diagram Figure 2. Here we show embeddings of an axially symmetrical cross-section of the metric given by (\ref{27}) and (\ref{29}). To display the curvature of a $ z=const$ cross-section of the space, we invent a new coordinate $u$ unrelated to $z$ and plot it upwards while $R$ is plotted outwards. We set $n^2C^2(R/b)^{2nm^2}dR^2+R^2d\B\vf^2=du^2+dR^2+R^2d\B\vf^2.$ So at constant $t$ and $z$, spatial lengths are measured along the $u(R)$ surface where $ (du/dR)^2=[n^2C^2(R/b)^{2nm^2}-1]$. In Figure 2 for $R/b>1$ we draw surfaces for no vertical stress, ($P_z=0;  C=1$), and for various values of $m$. Notice that  when $m=1$ the embedding is cylindrical with $R=b$ for all $u$, and all $C$. It is this independence of $g_{\vf\vf}$ that allows the transformation $\tt=i\B\vf$ and $\td\vf=i\B t$ which gives the metric its flat space form.
 Such cylindrical spaces have no internal curvature which explains why Levi-Civita's metric is "flat" not only when $m=0$ but also  when $m=1$. The dominant energy condition limits such $P_z=0$ equilibrium cylinders to $m\le1$ i.e. $\td \si\le1$. For $m>1$ we have therefore plotted limiting configurations with $P_\vf/\si=+1$ but in strong tension along $z$. These lie along the lower right hand boundary in Figure 3.
 
  \begin{figure}[htbp]
\begin{center}
\includegraphics[width=8cm]{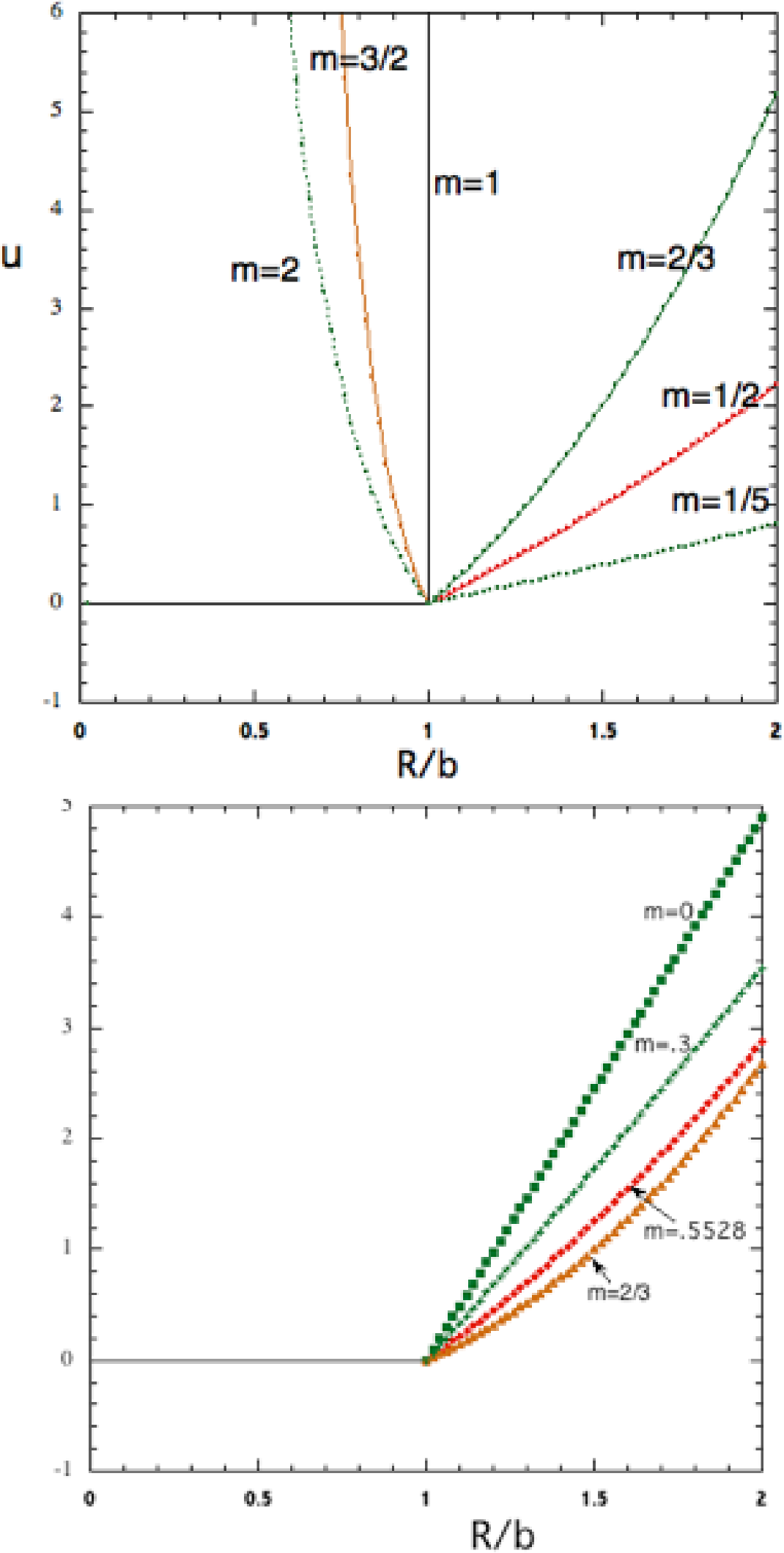}
\caption{ Embedding diagrams of $z=const$ cross-sections of  static cylindrical shells of radius $b$: (i) with no vertical pressure, Levi-Civita $m=1/5, 1/2, 2/3, 1$ i.e.     $2\pi b\si=\fr{2\pi}{\ka}(2m-m^2)$ when $C=1$. Increasing  the rest mass curls the embedding upward giving larger radial distance at a given circumference. For $m=1$ the embedding is itself cylindrical (for any $P_z$) and the space has no internal curvature. 
    For $m>1$ the shells disobey the dominant energy condition unless a vertical tension is introduced, so we set $P_\vf/\si=1$ and get for $P_/\si= - 2/3$ for $m=3/2$ and $P_z/\si= -1$ for $m=2$, the system at the bottom right corner of Figure 3.
    (ii) Embeddings of shells all of the same rest-mass  $2\pi b\si=\fr{0.72 \pi}{\ka}$ and various $C$. Vertical pressure decreases the upward curvature while vertical tension increases it having greater radial distance at given circumference (positive curvature). From bottom to top $P_z/\si=1, 0, -0.74, -1$.}
\label{fig2}
\end{center}
\end{figure}
We now ask what 3-force is needed to keep our test particle of rest mass $m_0$ in the motion\\
$u^\mu=\frac{1}{\sqrt{1-v^2}}\LLL\frac{\xi^\mu}{\xi}+v_\vf\frac{\eta^\mu}{\eta}+v_z\frac{Z^\mu}{Z}\RRR$?
Here $v^2=v_\vf^2+v_z^2$ , all those velocities are held fixed and $Z^\mu$ is the Killing vector along $\td z,~~Z=\sqrt{-Z^\mu Z_\mu}$. The 3-force required to keep the particle in this motion is,
\ba
f_k=\sqrt{1-v^2}m_0\ga_{kl}u^\nu D_\nu u^l\nn\\=\frac{m_0}{1-v^2}\partial_k[\ln\xi-v_\vf^2\ln\eta-v_z^2\ln Z] \nn\\
\lb{211}
f_R=\frac{m_0}{1-v^2}\LLL\frac{m}{(1-m)R}-\frac{v_\vf^2}{R}+\frac{mv_z^2}{R}\RRR
\lb{212}
\ea
The central term in the large bracket is the usual centripetal acceleration; the first is minus the radial gravity of the cylinder. This is always attractive towards the cylinder; for $m>1$ the apparent change of sign is due to the fact that the mass on the cylinder has so greatly bent the space that not only the flat inside but also the 'outside' has a smaller $R$ than the cylinder itself see Figure 2. Even when the gravity is pulling to greater $R$, it is still pulling towards the surface; the third is  an interesting effect that arrises because the gravity has curved (outwards) the lines with $R$ and $\B \vf$ constant that we think of as parallel to the axis. A particle travelling along one of these suffers a curvature acceleration. Thus in this term the gravitational bending of space and inertial acceleration combine to give the effect.   It is interesting to look at the equations of geodesics in these metrics. With a dot denoting $d/ds$, we have $\rho^{2nm}\dot t=E;\,\,R^2\dot \vf=h$
and $E^2=n^2C^2\rho^{2m}(\dot R^2+\rho^{2nm}(d\Bz/ds)^2)+\rho^{2nm}h^2/R^2+\rho^{2nm}$. Evidently
when $nm>1~i.e.~ 1>m>\2$ the centrifugal potential $\2(R/b)^{2nm}h^2/R^2$ at constant $h$ instead of decreasing to larger $R$, actually increases giving an effective net attraction toward the axis. This should not amaze us; the extra mass associated with extra transverse motion gives an excess gravitational attraction that can more than offset the centrifugal repulsion (Abramowitz 1993). As there is then nothing to offset gravity there are no circular geodesics for $m>\2$. This led Embacher (1983) to consider such systems as unphysical but they obey the dominant energy condition and, now the attraction is understood, we see no reason to reject them.The effect is also present in Schwarzschild's metric where orbits of constant angular momentum have a centrifugal potential $\2(1-2m/r)h^2/r^2$, so they are pulled into the origin for $r<3m$ as we saw in Section 2.2.
 \begin{figure}[htbp]
\begin{center}
\includegraphics[width=8cm]{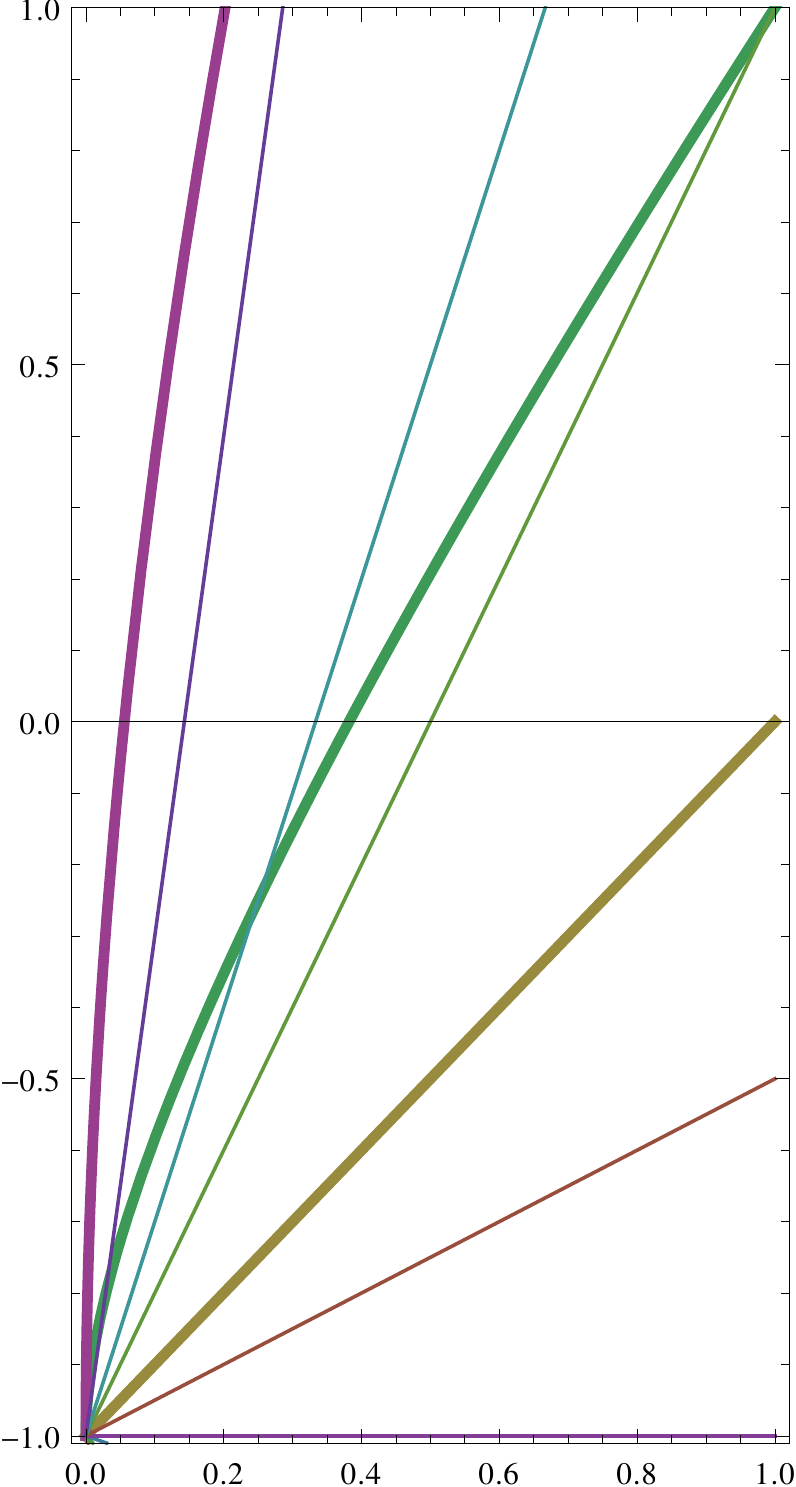}
\caption{Our $x$ is $P_\vf/\si$; our $y$ is $P_z/\si$; the dominant positive energy condition on our static cylindrical shells  is $0\le x\le1,-1\le y\le1$ and $\si \ge 0$. The straight lines emanating from (0,-1) have $m=2$ (along $y= -1$),$3/2,1,2/3,1/2,1/5$ and $m=0$ ($y$ axis). Heavy parabolae of constant rest mass per unit length $ 2\pi b\si=\fr{2\pi}{\ka}\LLL 1-(1-m)^2/C\RRR$ are superposed.  For $x<0$, $\si$ would be negative.}
\label{fig3}
\end{center}
\end{figure}

Since we are going to change $\Pi_z$, the longitudinal stress, it seems most natural to keep both the radius of the cylinder $b$ and its surface mass density $\si$ constant. Then $(1-m)^2/C=\De$ is constant. Systems with other radii can be found by rescaling. Evidently $\ka\Pi_z/(2\pi)=\De/(1-m)^2-1~,~\ka P_\vf=[m/(1-m)]^2\De/b$. When we increase $\Pi_z$, starting from zero, then $m,P_{\vf} $ will have to increase too to preserve equilibrium. Thus the gravitational field  strength $\EE=2 m/[(1-m)R]$ will increase at each $R$. Thus one effect of increased longitudinal pressure is an increased gravity for a fixed surface density $\sigma$. Another effect is illustrated in figure 2b in which embedding diagrams are drawn with the proper mass density on the shell held fixed; as the vertical pressure is increased the embedding diagram opens to become flatter but when the vertical pressure decreases to become a tension the embedding gets taller with  greater radial distance measured along the curve $u(R)$ corresponding to a given circumference $R$. However, even if  we can only measure properties outside the cylinder, we can still determine how much of Levi-Civita's mass parameter  $m$ arrises from the longitudinal stress because we can measure not only $m$ but also the rate at which the circumference of a circle changes with radial distance. This gives $2\pi dR/[nC(R/b)^{nm^2}dR]=[\De/(1-m)](b/R)^{m^2/(1-m)}$. With $R$ and $m$ known, observation of the cylinder's radius $b$ allows us to deduce $C$ from this. If that radius is not known then only $Cb^{-nm^2}$ can be deduced.

 Comparing two cylindrical shells of different vertical pressures and the same radius the extra pressure causes both an increase of gravity at a given $R$ and a larger value of $(1-m)/C$. When the longitudinal pressure is positive the rate of increase of circumference with radial distance measured at $R=b$ is  $2\pi(1-m)/C$ which increases. So increased vertical pressure produces more room at larger distance.
 
 When the cylindrical shell is in vertical tension the gravitating mass $m$ is reduced and $1/C<1$
Thus tension decreases the gravitating power although tension along one direction can not cause gravitational repulsion without violating causality via the dominant energy condition. With $1/C<1$ there is 
a lesser increase in circumference per unit increase in radial distance so the effect of tension is toward
a positive curvature of the space. Both these facts agree with what cosmologists already know about the Lambda term when that is treated as a form of matter. 

The possible equilibrium configurations are nicely illustrated in Figure 3  which plots $P_z/\si=y$ against $P_\vf/\si=x$. Since $x$ has to be positive to preserve equilibrium, the dominant energy conditions show that  the only possible configurations lie with $0\le x\le1$. The lines of constant $m$ radiate from $(0,-1)$, with $m=0$ the vertical axis, and $m=2$  the horizontal one. 
 
Equilibria with no $P_z$ i.e on $y=0$ and obeying the dominant energy conditions have $0\le m\le1$. For $m=1$ the space  is locally flat but with the identification of the planes $\td \vf=2N\pi$. In Figure 3 the curves  are the parabolae of constant rest mass per unit height, with  fixed $b$; mathematically these are given by 
$(1+y-x)^2=4(\td \si^{-1}-1)x$. Traversing such a parabola up from $P_z=0$, i.e. $y=0$, shows the change in $m$ and hence the change in gravity $m/(1-m)$, due to increased pressure is not impressive, however the traverse downward shows that when tension is close to the limit the gravity becomes very small. Below the line $y=x-1$ there is a region that can not be reached by following a line of constant $\td\si$ down from one of the  equilibria with $y=0$. This lower right corner of the diagram corresponds to $1\le m\le 2$ and in Figure 2 we see that even "outside" the cylinder $R$ is again smaller than $b$ as envisaged by Goldwirth \& Katz (1995). Our expression for the gravity $[m/(1-m)]/R$  is now negative, however as the "external" surface of the cylinder is now at larger $R$, gravity is still pulling towards that surface. Notice this effect sets in at $m=1$ and is distinct from the overwhelming of the centrifugal force which sets in at $m=\2$.
 \subsection{The Helical Gravity of Torque}
 
  In Figure 4 we draw ellipses oriented along the principal axes of the pressure (i.e. stress) tensor around a cylindrical shell subjected to both a longitudinal pressure and a torque. Notice that they are tipped at an angle to the axis. A little thought about the effect of the larger pressure along the longer axes of these tilted ellipses will convince the reader that they give a torque that carries angular momentum upwards. We engrave the usual $\vf,z$ coordinate lines on our shell and see that principal axes of stress are tilted with respect to them. The only static cylindrical empty solution of Einstein's equations that is regular on the axis is flat space. To fit to the coordinates engraved on the shell we can take the flat internal metric to be the usual $ds^2=dt^2-(dR^2 +R^2d\vf^2+dz^2)$. We must fit this 
 to the exterior metric on $R=b$ in such a way that we give a pressure tensor whose principal axes make helices on the shell. The space part of the pressure tensor has components $P_{RR},P_{R\vf},P_{\vf\vf}$. The time-time component is $\sigma$ and there is only one constraint equation, the radial component of the contracted Bianchi identity. Thus there are three independent components in the surface stress-energy tensor. To accommodate these the external metric needs three parameters corresponding to the mass per unit length, the momentum current up the cylindrical shell and the angular momentum current up it. In its most primitive form given above the Levi-Civita metric has only one parameter, m, but in the form given in equation (\ref{28}) it has two, $m,C$; now we see that a third is needed in its more general form. 
  \begin{figure}[htbp]
\begin{center}
\includegraphics[width=5cm]{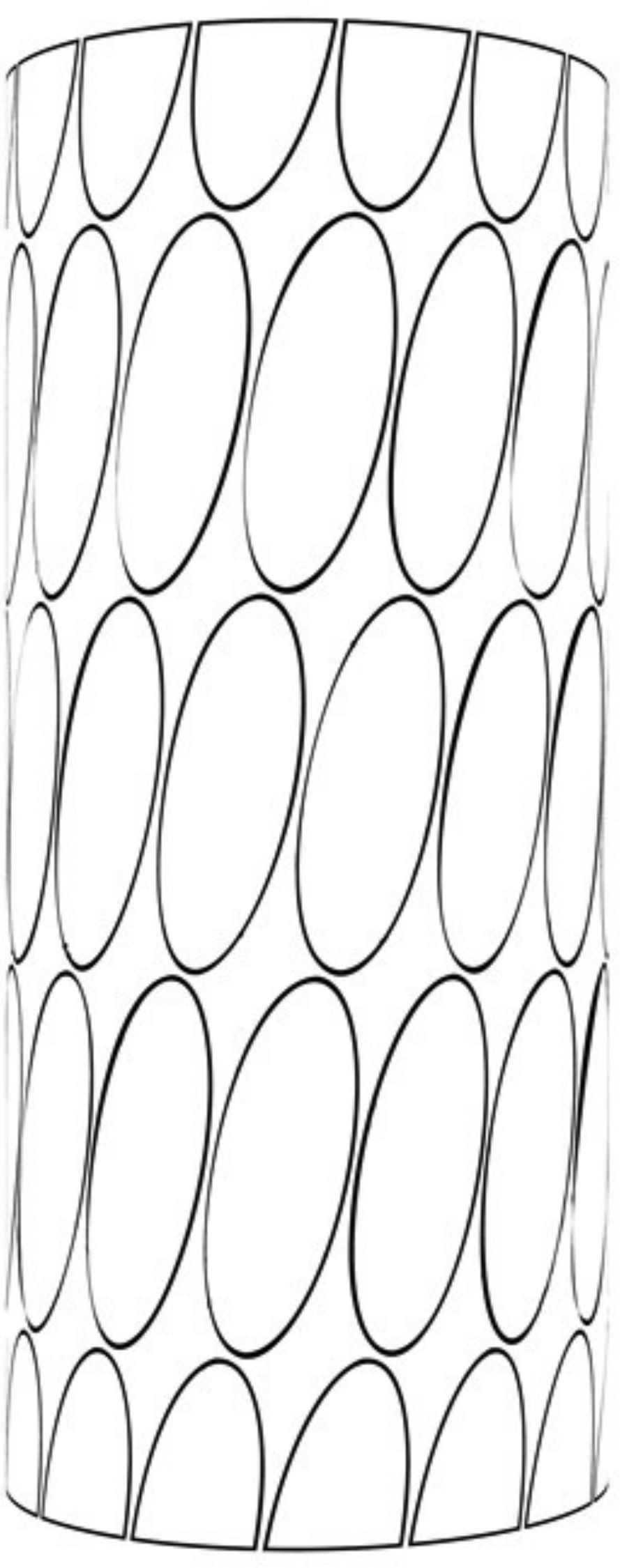}
\caption{The inclined stress-tensor ellipses of the torqued stressed cylindrical shell.}
\label{fig4}
\end{center}
\end{figure}
 The helices mentioned above do not have a $z\rightarrow-z$ symmetry but do have the $(z,\vf)\rightarrow(-z,-\vf)$ symmetry. 
 
 The empty metric with that symmetry can be found by plotting $\td z/b$ against $\B \vf$ and  making the transformation $~\B\vf=(\cos\al~\vf-\sin\al~z/b);~\td z/b=(\cos\al~z/b+\sin\al~\vf); $ in the Levi-Civita metric (\ref{26}). See Figure 5. On the cylinder $R=b$ this transformation is orthogonal.  With  $l_1=\rho^{-2m}\cos^2\al+\rho^2\sin^2\al$ and  $l_2=\rho^{-2m}\sin^2\al+\rho^2\cos^2\al$ the resulting metric is
 \ba
 ds^2=\rho^{2nm}d\B t^2-[ n^2 C^2 \rho^{2nm^2} dR^2+ l_2 b^2d\vf^2]-\nn\\
~~~~-[ b\sin2\al(\rho^{-2m}-\rho^2)d\vf dz+l_1dz^2].
 \lb{213}
\ea
This has three parameters $m,C,\al$ and reduces to the internal metric on the surface $R=b,\rho=1$ where $dR=0,~l_1=l_2=1$. It is this new $\vf$ that has the range $[0,2\pi)$.
Whereas $z$ and $\vf$ are orthogonal on the cylinder, at larger $R$ the cross term shows they are not. Nevertheless the gamma metric in the square brackets is everywhere positive definite.
We are now able to fit this external metric to the stress-energy tensor of a shell carrying both pressure and torque. Employing the variation of Israel's technique used earlier we find the stress energy tensor integrated across the shell is given by
\ba
\sigma=(\ka b)^{-1}[1-(1-m)^2/C];~~~~~~~\nn\\P_{\vf\vf}=\ka^{-1}(b/C)[m^2\cos^2\al+\sin^2\al]~~~~~~~~~~\nn\\
\lb{214}
P_{\vf z}=\ka^{-1}(1-m^2)bC^{-1}\sin\al \cos\al;~~~~~~~~~~\nn\\P_{zz}=(\ka b)^{-1}[C^{-1}\cos^2\al-1+m^2C^{-1}\sin^2\al]~~~
\lb{215}
\ea
These  agree with our former results and those of \cite{BZ} when there is no torque. 

The total flux of angular momentum up the cylinder is $\dot L=2\pi P_{\vf z}$ and the flux of momentum is now $2\pi b P_{zz}$.
It appears that all we have done is to change coordinates in  Levi-Civita's metric but that is a mathematician's viewpoint taking no regard to the interpretation and the ranges of the variables. Our physical cylinder could have the usual coordinates $\vf,z$ engraved upon it and these agree with the metric induced upon it from both the internal and the external metric. The external coordinates $\B t,R,\vf,z$ still have the Killing symmetries in $\B t,\vf,z$ but the specific angular momentum of a test particle in orbit is not given by $R^2 d\vf/ds$. Instead the specific angular momentum is $h=l_2d\vf/ds+b\sin\al\cos\al [\rho^{-2m}-\rho^2]dz/ds$ and the z-momentum is $p=l_1dz/ds+b\sin\al\cos\al[\rho^{-2m}-\rho^2]d\vf/ds$.  It is of course true that the dynamics is all much simpler if we return to Levi-Civita's metric in the form (2.7) with coordinates $\B\vf, \td z$ as in them we get back to our usual formulae $\B h=R^2d\B \vf/ds:~ \td p=\rho^{-2m}d\td z/ds$ but those coordinates wind in helices around our cylinder and $\B h,~\td p$ though conserved are not the specific angular momentum and momentum about and along the  axis of our cylinder. Nevertheless writing these quantities in our new coordinates $\B h=R^2[\cos\al~d\vf/ds-\sin\al~(dz/ds)/b]$;\\
 $\td p=\rho^{-2m}[\cos\al~ dz/ds+b\sin\al~ d\vf/ds]$. Solving these for $d\vf/ds,dz/ds$ we find,
\ba
d\vf/ds=\cos\al~\B h/R^2+\sin\al~\td p~\rho^{2m}/b;\nn\\~~~ dz/ds=\cos\al~\td p~\rho^{2m}-\sin\al~b\B h/R^2.
\lb{216}
\ea
Expressing the new linear and angular momenta in terms of the old,
\be
p=-(\B h/b)\sin\al+\td p\cos\al;~~~h=\B h\cos\al+b\td p\sin\al.
\lb{217}
\ee
   \begin{figure}[htbp]
\begin{center}
\includegraphics[width=8cm]{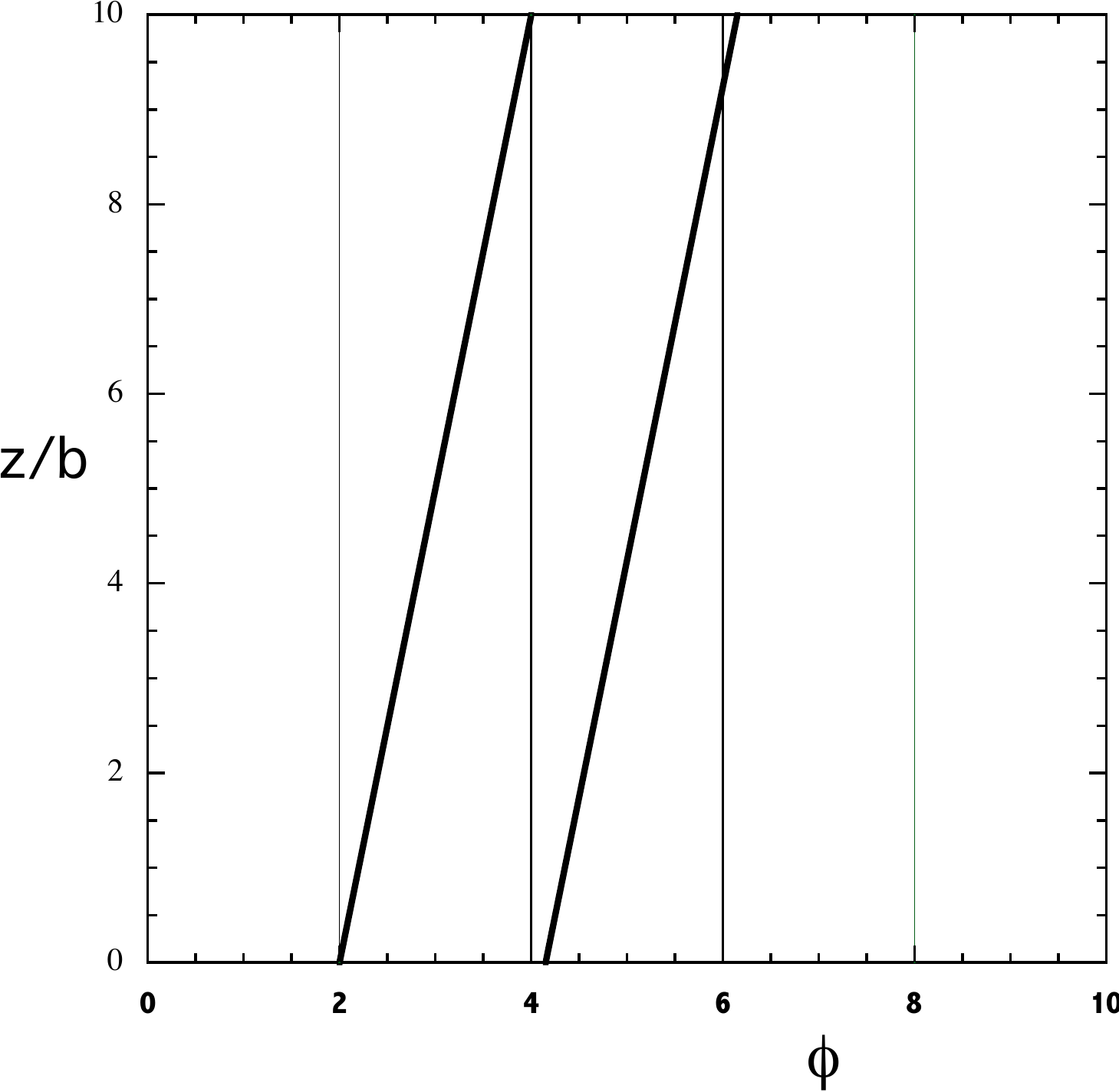}
\caption{Levi-Civita's coordinates $\B\vf,\td z/b$ lie at an angle in the covering space $\vf,z$ of the physical cylinder; units are multiples of $\pi$. The coordinate transformation is orthogonal.}
\label{fig5}
\end{center}
\end{figure}
 \begin{figure}[htbp]
\begin{center}
\includegraphics[width=8cm]{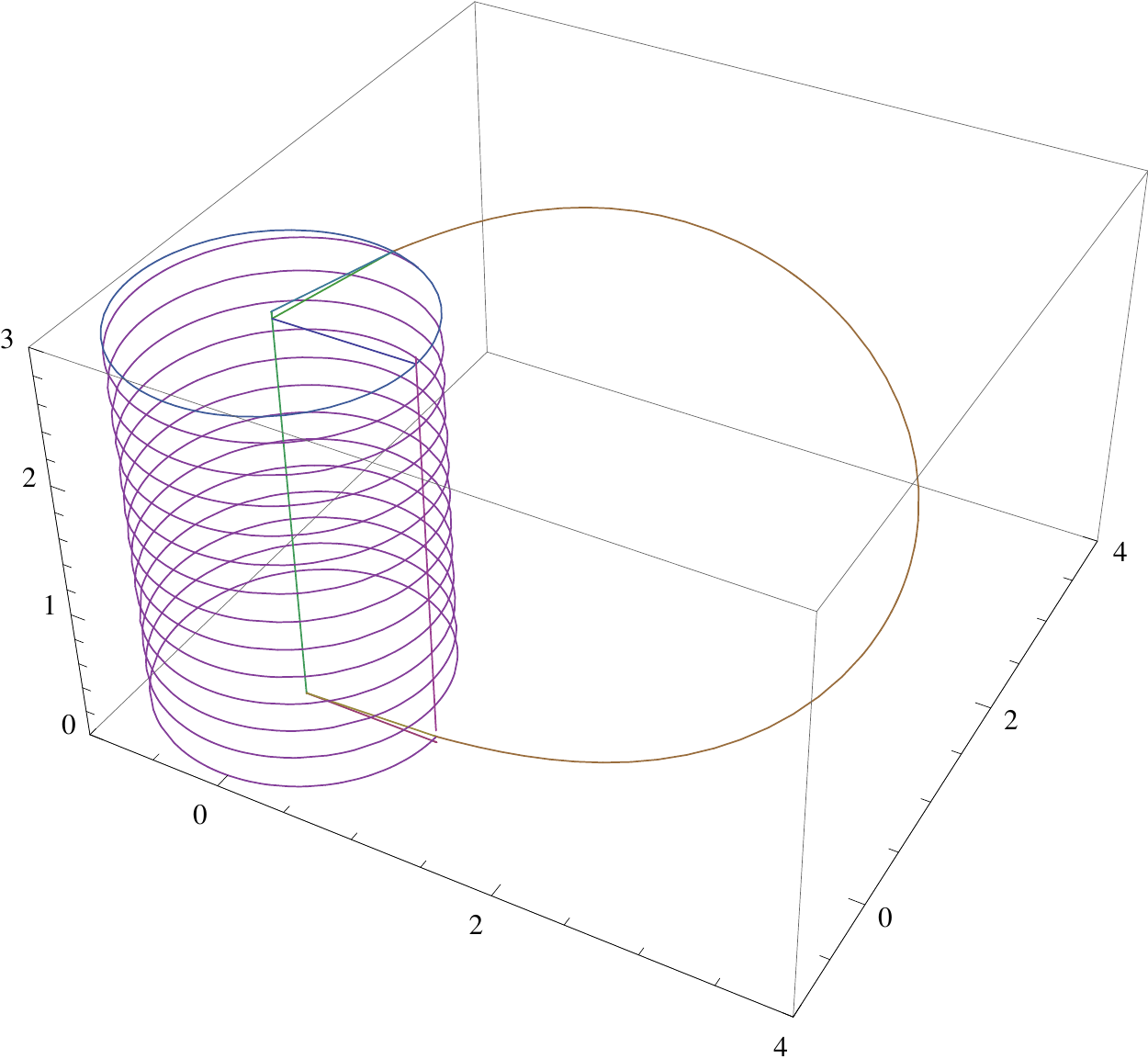}
\caption{ A geodesic with no angular momentum starts from the origin moving outward and upward in the 
xz-plane;  it emerges through the torqued cylinder still in that plane but the helical distortion of the space outside causes it to move in azimuth as well as radius and height. When it re-enters the cylinder it again moves in a plane through the axis. We chose the shear $\tan\al=0.5$ for this illustration.}
\label{fig6}
\end{center}
\end{figure}

What are the gravitational effects of a cylinder that carries torque? Firstly since $\xi$ depends only on $R$ its gravity is still radially directed to the axis. The geodesics inside the cylinder with zero angular momentum lie in planes $\vf=const$ but if we start a geodesic moving outward and upward
from the cylinder's surface then initially $d\vf/ds=0$ and from equation (\ref{216}) $\B h/b
=-\td p\tan\al$ so outside the cylinder it obeys $d\vf/ds=(\td p/b)\rho^{-2}(\rho^{2m+2}-1)\sin\al$,
which maintains its sign since outside $\rho>1$. When it again returns to the cylinder it will do so at another azimuth $\vf$ but at that moment it will again be moving in an $R,z$ plane since $\rho=1$. Thus the zero angular momentum geodesics wind around the cylinder whenever $\td p \ne 0$. The purely radial geodesics are the only ones that do not wind.
The forces on particles forced to move along the Killing vectors of this space are all radial since they can be put in the form (\ref{22}) and since all metric coefficients have only an R dependence the $\ze$ will have just that dependence. ( It will of course depend on the constants that define which Killing vector we are forcing our particle to move along.) However there are forces in the $\vf$ direction for trajectories that move in both $R$ and $z$.
The geodesics are fully integrable as is readily seen when we add the energy equation $\rho^{2nm}dt/ds=E$ to the integrals already found in (\ref{217}). Of course one must use the metric equation divided by $ds^2$ too. This cylindrical system has much to teach us about our intuition as well as its dynamics.\\
\subsection{Repulsive Gravity}
Is it possible to devise a static source of a vacuum field that obeys the energy condition 
$|p_i|\le \rho\ge 0$ and yet without any external Lambda term repels a static test mass?
This is not possible for spherical or stationary cylindrical sources however consider a plane circular sheet of mass-surface-density $\si$ under an isotropic tension per unit length $\si$. Then $ P_R=P_\phi=-\si<0; P_z=0$. Such a sheet on the $x,y $ plane generates gravity near the centre  $\EE_{|z|}=\4~\ka\si>0 $  directed away from the sheet. It is likely that such a finite sheet could be held in tension attached to a large ring. While astronomers already believe repulsive gravity is driving the acceleration of the universe it would be an amazing feat to demonstrate repulsive gravity in the laboratory! Just to find anything that can withstand tensions $P=-\si$ may prove impossible even before any attempt to measure its gravity. A straight magnetic field along $O_z$ has $p_z^z=-B^2/(8\pi)$ and energy density $B^2/(8\pi)$ but accompanied by unwanted positive lateral pressures of the same amount which destroy the effect, but even the gravity of magnetic field energy has so far escaped  laboratory detection, however see section 3.3.
	Relativity clearly shows that tension  reduces gravity and increases the spatial closure. It also
predicts that if substances or fields exist with $\rho+p_x^x+p_y^y+p_z^z<~0$ then they will show negative gravity and increase closure. Astronomers need such fields to get the right almost closed universe detected
via the CMB and independently to explain the observed acceleration of the universe. Thus they deem dark
energy to exist since both of its effects are seen. As yet no-one has explained theoretically the magnitude 
found from those observations.


\section {Thought Experiments with stationary metrics.}
\setcounter{equation}{0}
 \subsection{Experiments of Franklin, Oersted and Ampere}
{\color{black} The phenomenon now known as St Elmo's Fire in which pointed objects such as the tops of ships' masts glow when a thunderstorm is near, was known to the Greeks and was recorded in Roman times by Pliny. In de Bello Africo attributed to Julius Caesar  (circa 45 BC), it is reported that in February after a fierce hailstorm during the second watch the tips of the spears carried by the 5th legion glowed with a mysterious light. Later sailors noticed that these phenomena influenced the magnetic compass. However one of the earliest recorded suggestions that lightning is electrical is a letter to Dr Sloan secretary of the Royal Society written by a friend of Boyle's, Dr Wall in 1708. He describes his electrical experiments rubbing amber and diamond and states "Now I make no question but upon using a longer and larger piece of Amber both the cracklings and the light would be much greaterÉ and it seems to some degree to represent thunder and lightning." Some connection between such electrical phenomena and magnetism was reported in 1732 by Dr Cookson  of Wakefield Yorkshire. ``A tradesmen in this place having put a great number of knives and forks in the large box... and having placed the box in the corner of the large room, there happened a sudden storm of Thunder Lightning etc. by which that corner of the room was damaged, the box split and a good many knives and forks melted the sheaths being  untouched. The owner emptying the box upon a counter where nails lay, the persons who took up the knives that lay upon the nails observed that the knives took up the nails. Upon this the whole numbers were tried and found to do the same, nay to such a degree as to take up large nails packing needles and other iron things of considerable weight." In 1752 Benjamin Franklin' flew a kite as a thunderstorm approached and by charging a Leyden jar, definitively proved that lightning was electrical. The relationship of magnetism to electricity was only shown after Volta's discovery of the Voltaic pile battery in 1800. In July 1820 H.C.Oersted announced his discovery that a compass needle was deflected by a surrounding current loop to set perpendicularly to its plane. Oersted's discovery of the connection between electricity and magnetism was described by Arago at a meeting of the French Academy on September 11. Within a week Ampere showed that parallel currents attract each other and antiparallel currents repel. His experiments were not only beautifully devised and executed but they were also elegantly described}.\\
\subsection{Strong-Field Gravomagnetism}
Before we discuss the gravitational analogues of these experiments it is useful to put Einstein's equations
in a form appropriate for stationary metrics that are not static. The equations were developed by Landau and Lifshitz (1958) and later put in an elegant mathematical form by Geroch (1971). 
{\color{black} Somewhat different formulations of the resulting equations can be found in Thorne, Price \& Macdonald (1986), Lynden-Bell \& Nouri-Zonoz (1998), and Nata'rio (2005).} The metric can be written in the form
\be
ds^2=\xi^2(dt+\AAA_kdx^k)^2-\ga_{kl}dx^kdx^l
\lb{31}
\ee
 where $\xi^\mu\rightarrow(1,0,0,0)$ is the time-like Killing vector $\xi=\sqrt{\xi^\mu\xi_\mu}$ and the 3-vectors $\AAA_k$ and 3-tensors $\ga_{kl}$ are independent of time. We raise or lower their indices  by the spatial metric $\ga_{kl}$. The metric can be put in this form in more than one way. If we demand that $\xi$ become unity at infinity there is still the transformation of $t$ by a position dependent zero point $\B t=t-t_0(x,y,z)$ and $\B \AAA_k=\AAA_k+\partial_k t_0$. Then $\xi^2, \ga_{kl}$ are unchanged,  the form of metric is unchanged but ${\bf\AAA}$ undergoes a three dimensional gauge transformation. Since this is merely a coordinate transformation, all physical quantities are gauge invariant.  For stationary axially symmetrical systems there are two independent Killing vectors corresponding to shifts in time $\xi$ or angle about the axis $\eta$. Of course $a\xi+b\eta$ with $a,b$ constant are also Killing vectors. However with $b$ non-zero that Killing vector will cease to be time-like beyond a "light-cylinder". We therefore require that our time-like Killing vector becomes the usual $t$ at asymptotically flat infinity. As usual mass and angular momentum can be read from the asymptotic form of the matrix there. Cylindrical systems do not become asymptotically flat and need special treatment even in classical physics because the potential does not converge but behaves as $-2GM\ln(R/b)$. In relativity the space-time inside a hollow cylinder is flat even when the cylinder is under lengthwise pressure and is rotating and torqued. For non-singular solutions we follow the usual procedure of normalising the time-like Killing vector on the axis. Some authors (e.g. Embacher) have chosen to take the internal metric in the usual form for flat space $ds^2=dt^2-(dR^2+R^2d\vf^2+dz^2)$ and the corresponding time-like Killing vector, even when the cylinder rotates. While this has some advantages the resulting time-like Killing vector can not be continuous to the outside of the light cylinder. Thus it disobeys the criterion that the $t$ at large distances should not differ from the static one of Levi-Civita's metric. However that criterion is rather too restrictive. A cylindrical body surrounded by vacuum will in general have a mass, an angular momentum, a longitudinal pressure, a linear momentum along itself and carry a torque. All of these can be read out from the form of the metric at large radius from the coefficients of $dt^2, dtd\vf,d\vf^2,dt dz,d\vf dz,dR^2 ~{\rm and} ~dz^2$. It is remarkable that this five parameter family of solutions can all be found from simple coordinate transformations of Levi-Civita's original one-parameter metric. The reason behind this fecundity is that this metric has three Killing vectors giving displacements in $t,\vf,z$ but to the local mathematics these all have equal status so there is nothing to stop us reinterpreting the symbols so that the real $t$ at infinity is some linear combination of Levi-Civita's $t,\vf,z$ and the new $\vf$ some other linear combination of those three etc. When in what follows we talk about the gravomagnetic field or the relativist's twist vector we must recognise that these quantities depend on our choice of the time-like Killing vector from which they are constructed. Thus the gravomagnetic field of a cylinder moving along itself will be azimuthal on the outside while the same cylinder brought to rest via different coordinates is treated with a different time-like Killing vector which gives no gravomagnetism.  It is only for systems with high symmetry that stationary coordinates can be found in such variety with different time-like Killing vectors.\\ 
 Once a particular time-like Killing vector $\xi^\mu$ has been chosen Landau and Lifshitz then rewrite Einstein's equations as equations in gamma space. To write them we first define
 $({\bf curl\AAA})^k=\ga^{-\2}\ep^{klm}\partial_l\AAA_m;\\~~{\bf div\cd\!\!\BBB}=\ga^{-\2}\partial_k(\ga^{\2}\BBB^k);~~({\bf grad}\psi)_k=\partial_k \psi. $
Defining ${\bf\BBB}={\bf curl\AAA}$ we have ${\bf div\!\cd\!\BBB}=0$ so ${\bf\BBB}$ carries the gravomagnetic flux.
We shall also use the gravomagnetic field vector ${\bf\HH}=\xi^3{\bf \BBB}$ and the antisymmetric tensor 
$F_{\mu\nu}=D_{[\mu}\xi_{\nu]}$ which obeys $D^\nu F_{\mu\nu}=R_{\mu\nu}\xi^\nu$.
Re-expressing the Ricci tensor by use of Einstein's equations gives for the time component
\be
\xi {\bf div\cd grad}\xi+\2\xi^4\BBB^2=R_{00}=\ka(T_{00}-\2g_{00}T)
\lb{32}
\ee
and the space components give 
\be
({\bf curl \HH})^k=-2\ka\xi T_0^k= -2\ka J^k
\lb{33}
\ee
which bears a striking resemblance to Maxwell's equation for a steady current. Evidently the current ${\bf J}$
has no divergence. Our gravomagnetic field intensity 3-vector ${\bf\HH}$ is essentially the same as
the relativists' twist 4-vector $\om^\mu=\ep^{\mu\nu\si\tau}\xi_\nu D_\si \xi_\tau$ which has $\om_0=0$ and $\om^k=\HH^k$.    The space-space components of Einstein's equations become
\be
P^{kl}+ \2\xi^2(\ga^{kl}\BBB^2-\BBB^k \BBB^l)-\xi^{;k;l}/\xi=R^{kl}=\ka(T^{kl}-\2g^{kl}T)
\lb{34}
\ee
where $P^{kl}$ is the Ricci tensor formed from the 3-dimensional gamma-metric and semi-colons indicate
covariant derivatives in that metric.

Relativists rightly regard any two metrics that can be transformed into each other by coordinate transformation as physically equivalent. Merely different descriptions of the same entity. However when 
I ask a physicist for a description of the electromagnetic field of a charge he does not describe the field
tensor $F_{\mu\nu}$ in a general frame but instead puts the charge at rest and describes its electric
field. This is of course the simplest description since the magnetic field is zero in that frame. However when
the physicist  describes the electromagnetic field of a wire carrying a current he does not choose the frame in which there is no current up the wire and give the resultant electric field but gives the toroidal magnetic field $2I/R$. If we want to describe the full range of relativistic effects
it is useful to treat the metric of a column of mass in motion along itself as different from a static column
albeit their effects are all transformable into each other. These considerations at once tell us that the
gravomagnetic induction described above is not a property of the space-time itself but rather a property 
of the space-time together with a picked out time-like Killing vector. Of course when  such a vector is
unique the induction becomes a property of the space-time.
\subsection{Gravomagnetism of Energy Currents and Light}
We now apply these equations to the gravity of a cylinder that is moving along its axis. By the analogy between equation (\ref{33}) and electromagnetism we see that the current formed by the moving cylinder
will generate a gravomagnetic field which lies in circles about the axis and is the gradient of a scalar.
Hence we find that ${\bf\HH}=-(\ka/\pi)I{\bf grad} \vf$, where $I$ is the total energy current and $\vf$ is
 the azimuthal coordinate. Thus as expected the gravomagnetic field's intensity falls off as $1/R$. The minus sign arises because masses attract whereas like charges repel, and the magnitude is four times that found in electromagnetism, a difference that can be traced to the tensorial nature of gravity. The metric of an upward-moving cylindrical shell is readily obtained by Lorentz transformation along the axis. $z=\ga_L(\B z-V\B t),\\~t=\ga_L(\B t-V\B z);~~\ga_L^{-2}=1-V^2$. This leaves the internal metric unchanged but the external metric (\ref{27}) becomes
\ba
&&ds^2=\xi^2(dt+\AAA dz)^2-[n^2C^2\rho^{2nm^2}d R^2+R^2d\B\vf^2+Fdz^2];\nn\\
&&\xi^2=\ga_L^2(\rho^{2nm}-V^2\rho^{-2m})\nn\\
&&\AAA= \ga^2_LV(\rho^{2nm}-\rho^{-2m})/\xi^2;\nn\\
&&F=\xi^2\AAA^2+\ga_L^2(\rho^{-2m}-V^2\rho^{2nm}).\lb{36}
\ea
From this we see the gravity field is still radial with \\$\EE_R=-\frac{m}{(1-m)R}\LLL1+\frac{(2-m)V^2}{\rho^{2m+2nm}-V^2}\RRR$. Naively from Lorentz contraction and energy of motion one might have expected a factor $\ga_L^2$ times the static result, but the curved spatial metric complicates that idea.

We now repeat our experiments to find the force on moving masses. On a test mass constrained to move radially at a constant velocity ${\bf v}$ we find we have to supply a 3-force ${\bf f}=-\frac{m_0}{\sqrt{1-v^2}}({\bf\EE}+\xi {\bf v \times \BBB})$ and when instead we move the particle along at constant $R,\vf$ we have to supply a 3-force ${\bf f}=-\frac{m_0}{\sqrt{1-v^2}}({\bf\EE}+\xi {\bf v \times \BBB}+\2v_z^2{\bf grad} \ln F)$. 

It is the $\frac{m_0}{\sqrt{1-v^2}}\xi {\bf v \times \BBB}$ that is related to Oersted's result relating
gravomagnetism to gravity. Ampere's result is somewhat modified by the final inertial term but apart from that and the normal gravitation in the $\EE$ field, we see that as expected parallel energy currents attract with four times the strength that parallel electric currents repel and that anti-parallel energy currents attract. 
A rather beautiful illustration of these effects was given in Bonnor's (1969) discussion  of the gravity of a beam of
light which he treats as a column of null dust with $T_0^0=-T^z_z=-T_z^0=T_0^z=\ve$. In the metrics
\ba
ds^2=A(dt-dz)^2+dt^2-(dx^2+dy^2+dz^2)\nn\\=(1+A)\LLL dt-\frac{Adz}{1+A}\RRR^2-\LLL dx^2+dy^2+\frac{dz^2}{1+A}\RRR
\lb{37}
\ea
Einstein's equations reduce to $(\partial^2_x+\partial^2_y)A=2\ka\ve$. This equation is linear so solutions
for $A$ can be added. Thus we find the remarkable fact that parallely propagating  beams of light do not affect each other gravitationally. Inside a cylindrical column of light of radius $b$ Bonnor has $A= 4mR^2/b^2$ and outside $A=4m+8m\ln(R/b)$. The corresponding gravomagnetic field is toroidal  $\HH_\vf=-2A/R,~~ R\le b;$ and $=-(8m/R)~~R\ge b.$ The corresponding gravomagnetic inductions $\BBB_\vf$ have an extra factor $(1+A)^{-3/2}$. Putting the   gravitational force $M=mc^2/G$ on a static test body due to such a column of light is  approximately $4Gm_0M/R$ and the total momentum transfer in time $\De t$ is multiplied by that factor. The momentum of light that has passed in that time is $Mc (c\De t)$ so the light will be bent by an angle $4GMc^{-2}/R$. This is the standard formula, a factor two greater than Cavendish's Newtonian one, but now derived from the gravity of the light! Whereas in the usual derivation half the effect comes from 
the potential and the other half from the purely spatial curvature of Schwarzschild's metric, here all the 
effect comes from the potential of the light beam coupled with momentum conservation.
Since a beam of light attracts a static test mass twice as strongly  as we might expect from its energy density, how is it that two such beams do not attract each other? The 3-force on a test mass moving at constant $R,\vf$ is minus the force we have to give to keep it so moving, so as  above $-{\bf f}=\frac{m_0}{\sqrt{1-v^2}}({\bf\EE}+\xi {\bf v \times \BBB}+\2v_z^2{\bf grad} \ln F)$, but now ${\bf\EE=-grad}\ln\xi;~~ \xi=\sqrt{1+A};~~~\\ \xi{\bf v\times\BBB}=v(1+A){\bf grad}[A/(1+A)];~~~\ln F=-\ln(1+A)$ thus  $-{\bf f}=-\frac{m_0}{\sqrt{1-v^2}}\2 (1-v)^2{\bf grad}(1+A)$. As expected from Bonnor's result this is zero at $v=1$. However as he pointed out
for opposed light beams passing one another, with $\frac{m_0}{\sqrt{1-v^2}}$ held fixed at a finite value, the force is four times that on a static body of the same mass i.e. 8 times what one would estimate from the energy density of the light beam. Also light beams passing each other perpendicularly will each bend in the gravity field of the other. Nearby this is just twice as much as they would bend in the presence of rods of the same energy density, but actually neither beam can escape from the other assuming both are infinitely long because $A$ diverges for large $R$!\\
	Another illustration that parallel currents of energy repel whereas antiparallel currents attract is found
in the equatorial orbits in Kerr's metric for a rotating black hole. A particle in a retrograde orbit feels this extra attraction which leads  it to fall into the hole from a larger distance than its companion that orbits with angular momentum parallel to the hole's spin. The latter particle is subject to the repulsion of parallel currents, since it moves with the hole's current, as well as the straight gravity of the hole so its last stable orbit is closer to the hole.
\subsection{Gravomagnetic Solenoids, Rotating Cylindrical Shells}
In electricity a long but finite solenoid with radius $b$ and length $2a$, carrying an electric current $I$ per unit length, has an internal field of $4\pi I$ and an internal magnetic flux  $F=4\pi^2b^2I$  which is also $4\pi$ times the pole strength at the ends. Outside the solenoid  the field due to those poles is $\frac{F}{4\pi}[{\bf r}_N/r_N^{-3}-{\bf r}_S/r_S^{-3}]$ where ${\bf r}_N$ is the vector measured from the North Pole of the solenoid and  ${\bf r}_S$ is from the South Pole. On the equatorial plane the external field is vertical and of strength $\frac{Fa}{2\pi(R^2+a^2)^{3/2}}$ which decreases like $a^{-2}$ as $a$ becomes very large. Nevertheless all the flux returns to the other end as $\int(R^2+a^2)^{-3/2}aRdR=1$. Although the magnetic field outside tends to 0 for large $a$ the total external flux remains the same but is spread over a wide area.
	In relativity the closest analogous exact solution is that for an infinite massive cylindrical shell rotating with angular velocity $\Om$ (Frehland 1971,1972., Embacher 1983) for which the metric is,
\ba &&ds^2=dt^2-[dR^2+R^2(d\vf-\om dt)^2+dz^2],~~\nn\\
&&~~~~=\xi^2(dt+\AAA d\vf)^2-\xi^{-2}[e^{2k}(dR^2+dz^2)+R^2d\vf^2,\nn\\
&&\AAA= \om R^2/\xi^2,~R<b,\nn; ~~\xi^2=1-\om^2R^2 =e^{2k},~R<b,\nn\\
&&ds^2=\xi^2(dt+\AAA d\vf)^2- \xi^{-2}[e^{2k}(dR^2+d\Bz^2)+R^2d\vf^2],\nn\\
&&\xi^2=(1-\om^2b^2)(R/b)^{2m},~R>b,~~\nn\\
&&\AAA=\frac{\om b^2}{1-\om^2b^2};~e^{2k}=C^2(R/b)^{-2m^2},~R>b.
\lb{311}
\ea
By putting the metric both inside and outside the shell in Weyl form we have been forced to make the coordinate $z$ discontinuous. To regain continuity we put $\Bz=z~\sqrt{1-\om^2b^2}/C$ outside.
From this metric we find the gravomagnetic field ${\bf\HH}=-2\om {\bf \hat{z}},~ R<b; {\bf\HH}=0,~R>b.$ A result directly comparable to the electromagnetic case. The gravomagnetic induction is ${\bf\BBB=\HH}/\xi^3$.
The surface-density of rest mass $\si$ is given by $\ka b\si=\td\si=2\td\mu/(1+y-x-2\td\mu y)$ where $x,y$ are the dimensionless ratios, $P_\vf/\si,P_z/\si$ defined earlier and $\td\mu=m(1-m)$. These equilibria depend on those three dimensionless numbers. The rest mass per unit length is $M_1=2\pi b\si$. The velocity $V$ of the cylinder relative to the rotating frame in the flat space inside  it is given by 
\be
V^2=\frac{\td\mu(1+x+y)^2-2x(1+y-x)}{\td\mu(1+x-y)^2+2(1+y-x)}.
\lb{312}
\ee
 The angular velocity of that frame is given by
\ba
 \om=\frac{1}{b}\sqrt{\frac{mQ_1-\td\mu Q_2}{(1-m)Q_1-\td\mu Q_2}};~~~~~~~~~~~~~~~~~~~~~~~\nn\\ 
 Q_1 =(1+y-x)(1+y-x-2\td\mu y); ~~~~~~~~~~~~~~~~~~~~~\nn\\Q_2=\td\mu (1+y+x)^2-2x(y-x)+1+y-x.
 \lb{313}~~~~~~~~~
 \ea
The Komar mass per unit height is $\2m/C$ where $C=1-[2\td\mu y/(1+y-x)]$. The  total momentum flux up the cylinder is $2\pi b P_z=y\td\si$. 
The angular velocity of the cylinder is $\Om=\fr{V}{b}+\om$.
Rotation increases the external gravity and decreases the azimuthal pressure.
 \begin{figure}[htbp]
\begin{center}
\includegraphics[width=4cm]{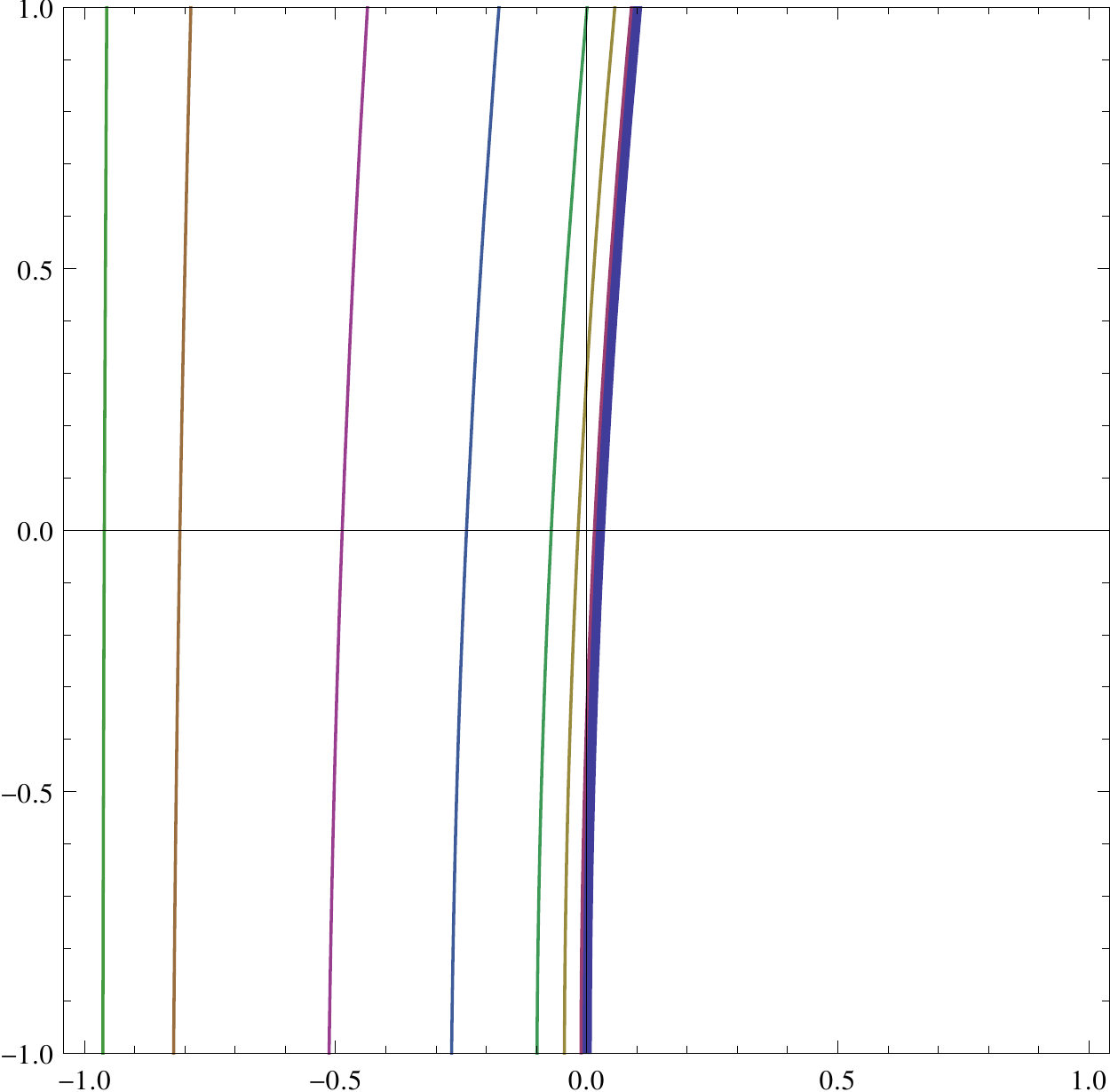}
\includegraphics[width=4cm]{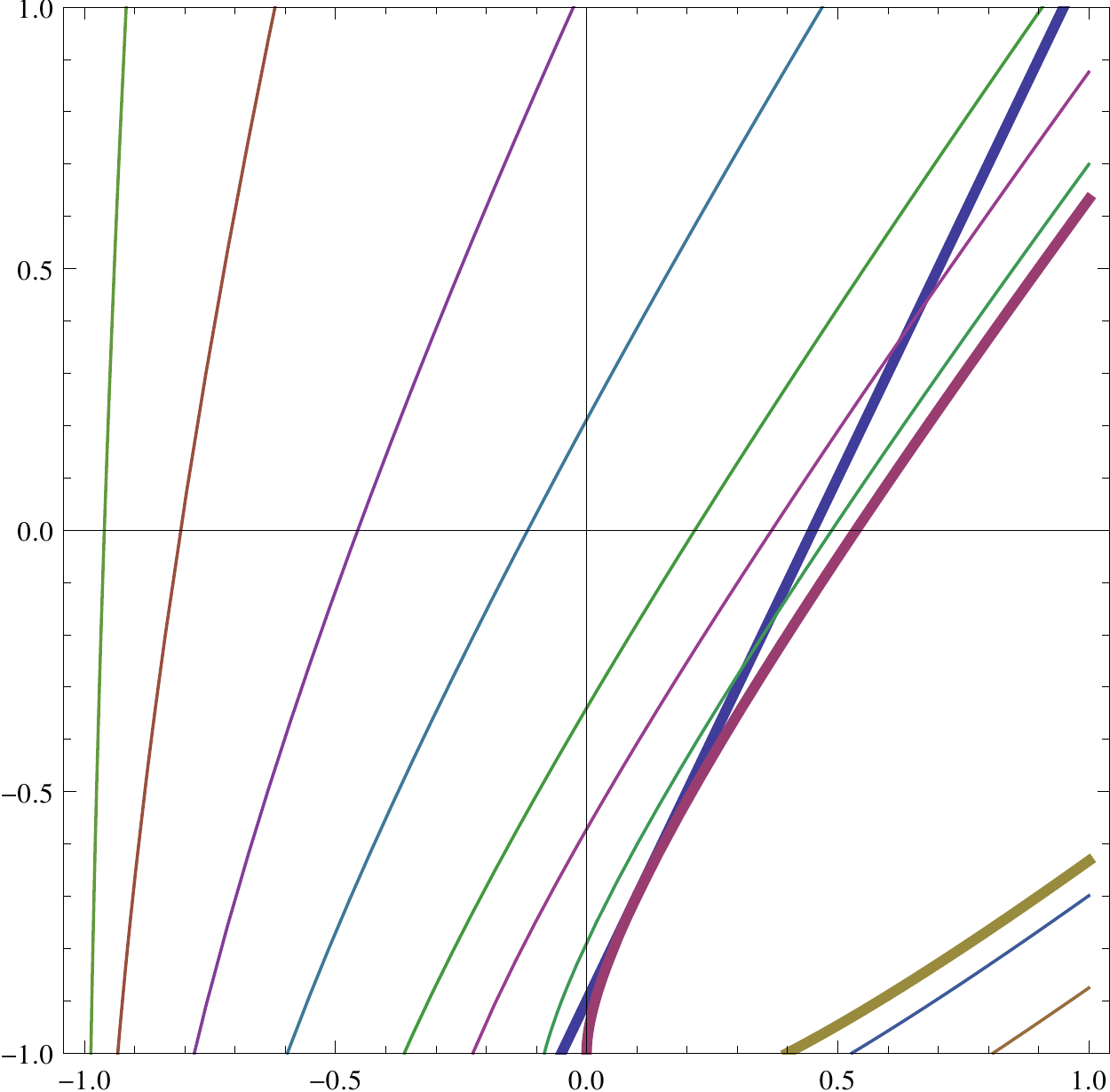}
\includegraphics[width=4cm]{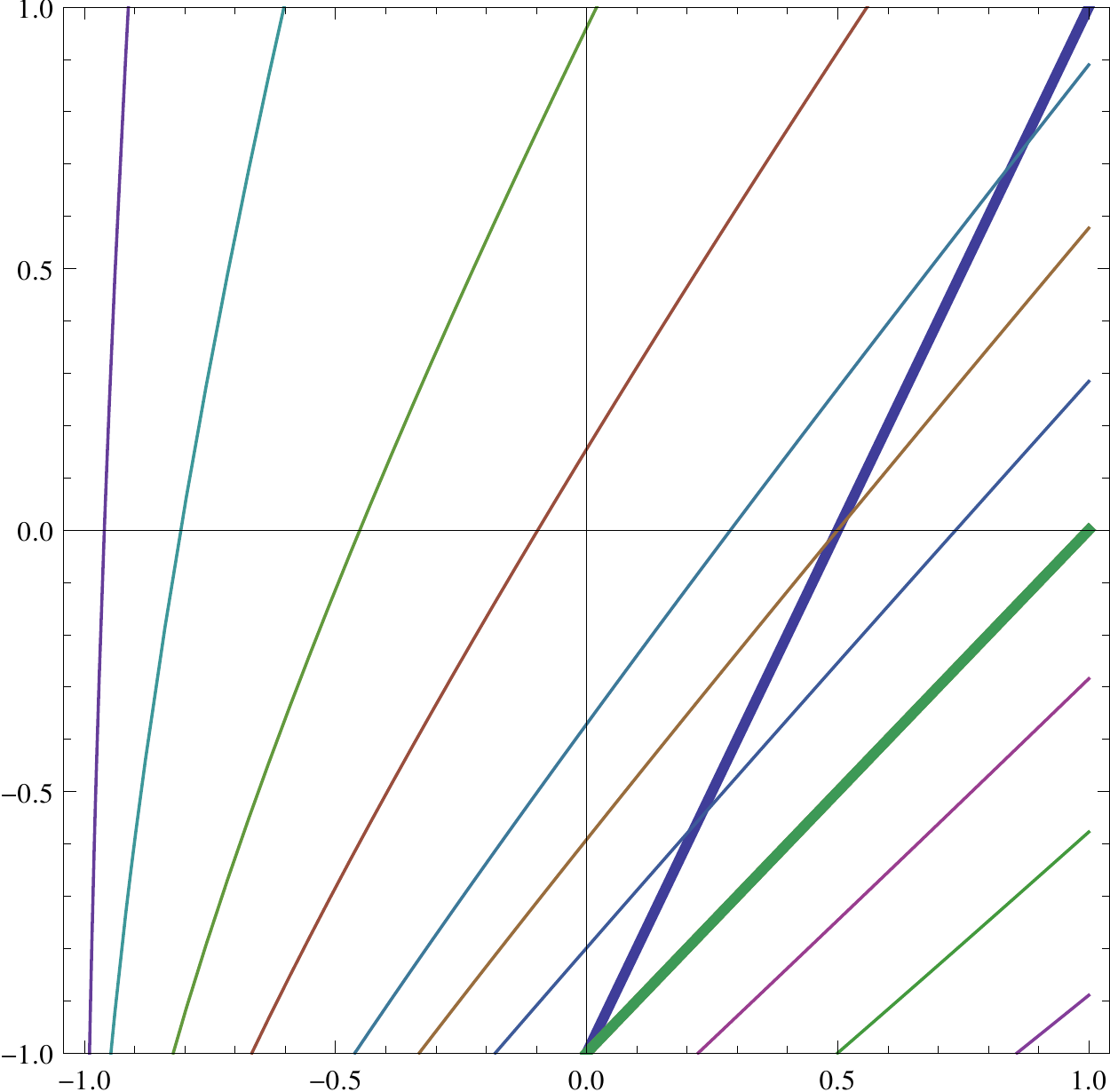}
\includegraphics[width=4cm]{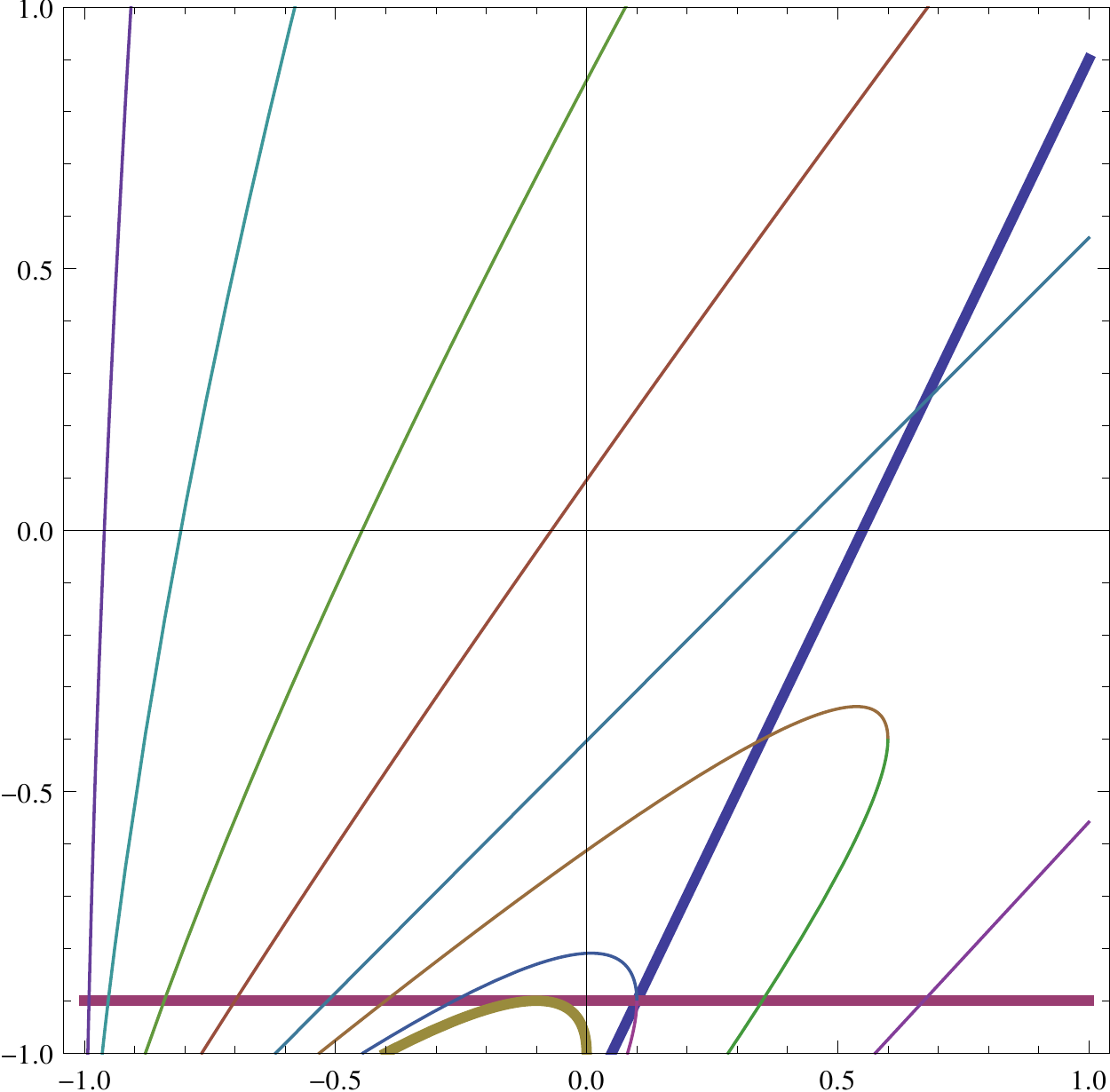}
\includegraphics[width=4cm]{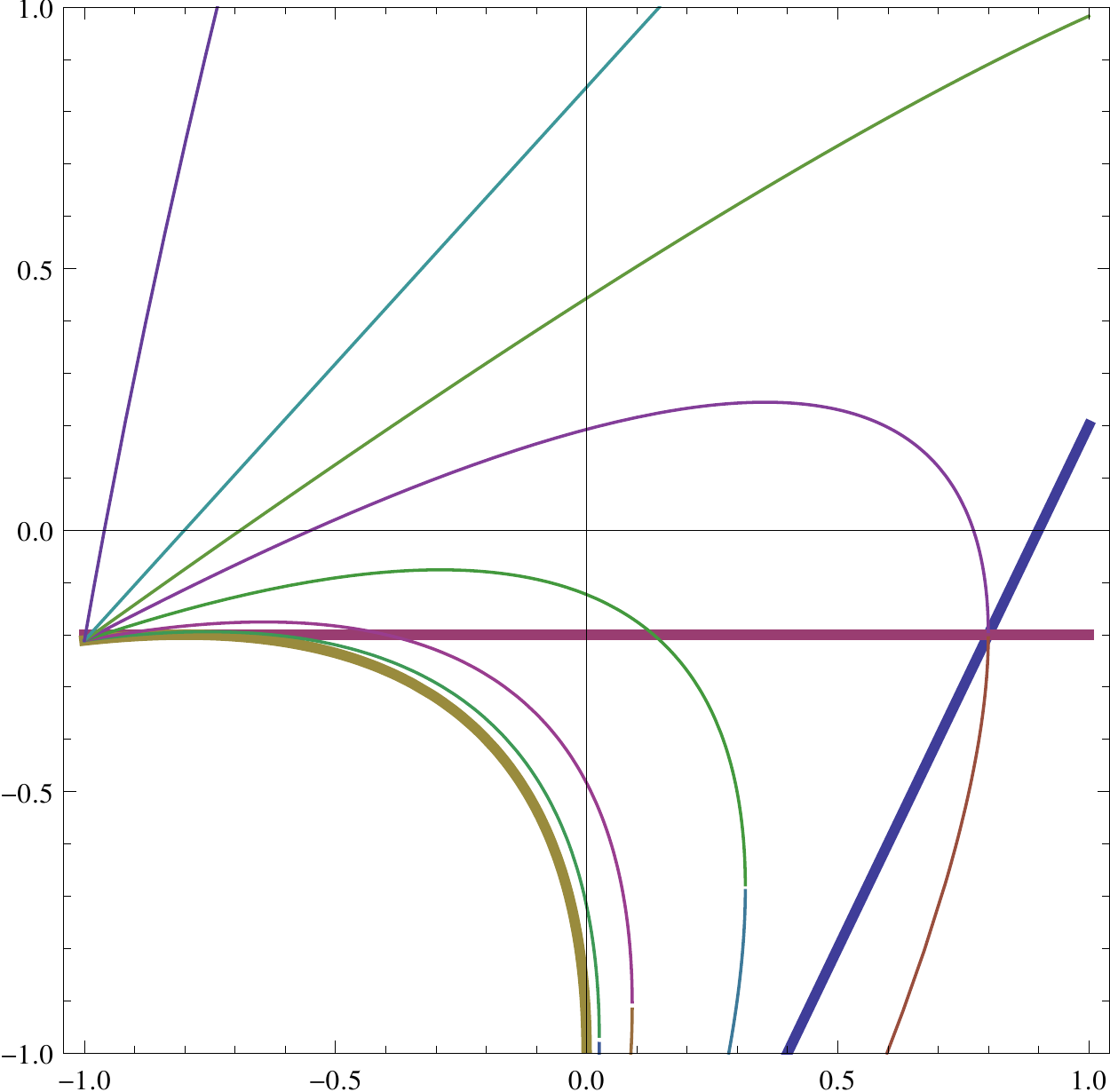}
\caption{  Rotating cylinders. x and y as in Figure 3. We plot lines of constant velocity of the cylinder with respect to its internal flat space which itself rotates relative to fixed axes at infinity. Each diagram is for a different  mass per unit length $2\pi b\si=2\pi\tsi/\ka$. $~~~~$
  (i) $\td \si=0.1$;$~~V/c=0 (heavy), .2, .3, .5,.7, .9, .98$ (ii) $\td\si=10/11$; $~V/c= 0$(heavy parabola),$ .1, .2, .3, .5, .7, .9, .98$ Only systems to the right of the heavy straight line are real.
  (iii) $\td\si=1;V/c=0$ (heavy through $(1,0)), .1, .2, .3, .5, .7, .9, .98$ 
  only systems below the line joining $(0,-1)$ and $(1,1)$ are real.
  (iv) $\td\si=10/9;V/c=0, .1, .2, .3, .5, .7, .9, .98$. Real systems between and to the upper right of heavy lines.
  (v) $\td\si=5;V/c=0, .3, .5, .7 .8, .85, .9, .98.$ Real systems as in (iv)}
\end{center}
\end{figure}
Externally $\AAA$ is constant so it can be eliminated locally by the transformation $\B t=t-\AAA\vf$
however $\B t$ is discontinuous (or multivalued) at $\vf=2N\pi$. Indeed as Stachel (1982) points out  
the interference of light passing each side of such a rotating cylinder is affected by the rotation even though there is no local effect on the space-time where the light passes. He points out the strong analogy with the quantum mechanical Aharanov-Bohm (1959) effect in magnetism for which see also Dirac (1931).

The full panoply of the rotating cylindrical shell solutions illustrated in Figure 7 show the relativistic force
of stress. In Figure 7(i) the surface density of matter is low so Newtonian ideas apply. The dominant energy condition ensures that the pressures and tensions are so weak that their gravitational effects are small. Any positive azimuthal pressure is limited by that which balances the weak gravity so no equilibria exist to the right if the heavy $V=0$ gently curved line. The role of the velocity is to offset some of the gravity via centrifugal force and at larger velocities to require a significant azimuthal tension to hold the cylinder together against the strain of centrifugal force. The vertical pressure along the cylinder has a small effect on the equilibria
via the small gravity increase that it produces but the diagram is nearly independent of $P_z/\si=y$. This is
clearly not the case in Figure 7(ii) which is plotted for a surface density of 10/11ths of what a static shell can withstand at equilibrium. Here an increase in vertical pressure gives an increase in gravity that must be opposed by the combined effects of centrifugal force and azimuthal pressure support. Again there is a boundary on the right where V=0 and all the gravity is opposed solely by azimuthal pressure, cf  Figure 3,
but now there is another heavy straight line to the left of it. This boundary arrises from the requirement that the "external'' empty-space solution is not so strongly dragged around by the rotation that it has closed time-like lines. van Stockum (1937) gave explicit real solutions of Einstein's equations
which correspond to imaginary $m-\2$ and inhabit the region to the left of this new boundary. But such  solutions have zeroes in their metric coefficients giving rise to closed time-like lines, as Tipler has emphasised, and do not obey the usual cylindrical boundary conditions at large $\B R$. Tipler (1974) demonstrated that there were no other solutions in this regime so we conclude that there are no acceptable external solutions to the left of the straight boundary. Thus acceptable solutions obeying the energy conditions lie between the heavy line and the heavy parabola $V=0$. Near the bottom right corner of Figure 7(ii) there are solutions with strongly negative $P_z$ corresponding to the static solutions in that region with $m>1$ mentioned earlier. In this regime the cylinder's surface at $\B R/b=R/b=1$ is the radius with the greatest circumference in all the space (see figure 2). The gravity in the region $\B R>b$ "outside" the cylinder acts toward larger $R$ but that is toward the cylinder. The gravitational attraction and the centrifugal force both
act outward toward the surface so for these solutions any increase in velocity must be accompanied by a decrease in gravity which can be achieved by an increase in vertical tension. Thus the run of velocity with
$y$ is in the opposite direction in these solutions as compared with those with $y>0$. Figure 7(iii) is for the special case $\td\si=1$. Here the upper and lower $V=0$ boundaries have coincided in the line $y+1=x$. Thus $V$ increases in both directions away from this line. The condition for no external time-like lines is now that both $y+1\le 2x$ and $y+1\ge 0$ but the latter is already ensured by the energy condition. As Bi\v{c}\'{a}k, \& \v{Z}ofka (2002) have emphasised, there are no real static solutions with $\td\si>1$  so the $V=0$ boundary in Figure 7(iv) which is drawn for the high surface density corresponding to $\td\si=10/9$ only reappears in the lower left of the diagram which is forbidden due to closed time-like lines. Acceptable solutions start at V=0.1 and occupy the right-angled triangle in the right of the diagram. Finally figure 7(v)
is drawn for the very high surface density  corresponding to $\td\si=5$. Here only a tiny triangle near the middle of the right side of the diagram gives acceptable solutions all of which have $V\ge 0.8$.

In electromagnetism a toroidal solenoid has its magnetic field confined inside the torus. The analogous gravitational solenoid is a toroidal shell with rolling motions around its small cross section, the inner equator moving up and the outer equator moving down (say). The absence of any external gravomagnetic field allowed us to use Weyl's technique to solve for the gravitational field there (Lynden-Bell \& Katz 2012) however the
gravitational Stachel-Aharanov-Bohm effect will still affect the interference of light some of which passes through the torus and some outside. Thus the flux of $\BBB$ through the small cross section of the torus
is detectable from its outside.
\section{Faraday, Lines of force and Induction}
\setcounter{equation}{0}
Faraday's picture of the electromagnetism was in terms of electric and magnetic lines of force and their fluxes. In relativity these are combined into the antisymmetric electromagnetic field tensor whose amazing flux carrying properties, which extend to curved space-times, have not been popularly expounded. This field tensor can be written in terms of the vector  potential ${\bf A}$ via $F_{\mu\nu}=\partial_\mu A_\nu-\partial_\nu A_\mu$ and is related to the 4-current $j^\mu$ in the form $D_\nu F^{\mu\nu}= 4\pi j^{\mu}$. Since $F^{\mu\nu}$ is antisymmetric $j^\mu$ is automatically conserved. Now any conserved electrical current has an associated electromagnetic field which can be described by a vector potential ${\bf A}$ which may be taken to obey the Lorenz condition $\partial_\mu A^\mu=0$ so by analogy any conserved current may be expressed as the divergence of a skew tensor which can be written as the antisymmetric derivative of a solenoidal vector field. Although general expressions for the energy current, the momentum current, and the angular momentum current have not been found, nevertheless total energy, total momentum and total angular momentum are conserved and Komar (1962) has given expressions for them in terms of the asymptotic Killing vectors. In the special cases in which a relevant Killing vector exists everywhere (i.e. not just asymptotically) Komar's expressions give fluxes that relate the totals calculated at infinity to their sources $T_{\mu\nu}$. Such relationships were found earlier in coordinate form by Tolman (1934) and by Whittaker (1935).

	Since ${\bf div\BBB}=0$ lines of gravomagnetic flux do not end and we can define them for all stationary spaces once a suitable frame of reference at infinity is chosen. In cylindrical systems this last requirement is particularly necessary as transformations to both rotating coordinates and those in motion along the axis still leave the metric in the stationary form (\ref{31}). To remove the rotational ambiguity we shall always require that the stationary frame  be chosen so that it does not rotate far from the axis. Thus our gravomagnetic field is defined once we fix the motion of our frame along the axis. The most natural frame to choose is one in which the system has no momentum along the axis, but as in electrical problems we may wish to use other frames to see the effects of energy currents etc.
	
	In our consideration of gravitational forces we have already introduced the usual gravity field $\EE= - {\bf grad}(\ln\xi)$. Both ${\bf\EE}$ and ${\bf\BBB}$ are defined in the gamma metric of space. This is not normally a cross section of space-time by a surface, but is rather the quotient space found by identifying the points along the time-like Killing vector ${\bf \xi}$. Thus an essential part of the definition of ${\bf \EE}$ and ${\bf\BBB}$ is the existence of such a Killing field. Indeed it is the non-local structure of this field that allows us to bring in from infinity the frame of reference used there. \\Faraday used fluxes of both electric field ${\bf E}$ and electric displacement ${\bf D}$. From equations (\ref{32}) and (\ref{33}) we find using,
\ba
{\bf div(\AAA\ti\HH)=(\bf curl\AAA)\cd\!\HH-\AAA\cd\bf curl\HH}~~~~~~\nn\\={\bf \BBB\cd\!\HH}+2\ka\xi\AAA_kT_0^k~~~~~~~~~~~~~~
\lb{41}
\\
{\bf div(grad \xi+\2 \AAA \times \HH) } =\ka(T_{00}/\xi-\2\xi T+\AAA_kT^k_0\xi)\nn\\=\ka(T^0_0-\2T)\xi~~~~~~~~~~~~~~~~~~\lb{42}
\ea
In the final equality we used $T_{00}=g_{0\mu}T_0^\mu=\xi^2(T_0^0-\AAA_kT_0^k)$.
The relationship of Faraday's fluxes of ${\bf \DD}=\xi{\bf \EE}$ and ${\bf \BBB}$ to the Tolman and Komar formulae is found by integrating the above over all space $\sqrt{\ga}d^3 x=dV$. The expression on the left
of (\ref{42}) converts to a flux  over the sphere at infinity where the magnetic term vanishes and for asymptotically flat spaces $\int {-\bf \DD.dS}$ evaluated there gives $4\pi G M$, since $\xi \rightarrow 1-m/r$. Thus we get Tolman's formula $4\pi GM=\int (T^0_0-\2T)\xi dV.$ Similarly from (\ref{41}) we find
\be
\int {\bf J\cd\!\AAA}dV=-\2\int {\bf\BBB\cd\!\HH}dV/\ka=\int\AAA_kT_0^k\xi dV
\lb{43}
\ee
This we regard as an equation for the magnetic energy of the configuration which is negative (remember that parallel currents attract in electricity but repel in gravity). 
\section{A Conundrum on the Definition of Gravitational Forces}
\setcounter{equation}{0}
 One of the greatest triumphs of general relativity is the unification of the laws of gravity and inertia. Newton's ideal space and time through which free particles travelled along straight lines at uniform speed, from which they could be deviated by gravitational forces, was replaced by the law that a particle moving freely under gravity follows the locally straightest path through the space-time in which it finds itself. Because space-time is not flat these locally straightest or geodesic paths are far from being globally straight but can be circles, near- ellipses or almost hyperbolae. In the freely falling frame of a particle gravity is locally eliminated so it became inappropriate to talk about gravitational forces and gravitational fields of force, however tidal forces on bodies can be expressed locally and tensorially so that these force differences that distort through the equation of geodesic deviation are quite widely used by relativists.\\
	To the engineer or architect who has to design a building which will stand against winds, hails, floods, and earth-tremors the idea that gravity can be eliminated by taking a freely falling frame of reference is irrelevant and unhelpful. The force of gravity is all too real and anyone designing a bridge had better study Newton rather than Einstein in order to calculate the stresses within the structure. Many astronomers and physicists who study dynamics have developed a good intuition as to how bodies or even orbits respond to extra forces but that intuition counts for little when they are given only a curved space-time to consider. Thus, in the early years, relativity was developed by geometers and the more theoretical astronomers and physicists with only a few  highly symmetrical exact strong field solutions becoming known. One of the biggest obstacles to mechanical intuition on how strongly relativistic systems work lies in the lack of understanding of gravitational forces and even in the lack of a suitable framework for defining them. Our aim here has been to give precise definitions of gravitational force fields in those special cases where we already have no ambiguity, in the hope that these may later be extended to cover more general situations.
	Landau and Lifshitz pursued this path and gave general expressions for the gravitational forces on
a particle moving in any stationary space-time. Our results above are special cases of theirs but we now expose a conundrum that arises from their more general definition of gravitational forces.
	 If a particle is left free, that is subject only to gravity, it pursues  a geodesic path through space-time. 
Landau and Lifshitz ascribe  the difference of that path from a geodesic of the spatial metric $\ga_{kl}$ as due to gravitational forces. Thus they write,
 \ba
 \sqrt{1-v^2}\frac{d}{ds}\LLL\frac{mv^k}{\sqrt{1-v^2}}\RRR+\la^k_{lm}\frac{mv^lv^m}{\sqrt{1-v^2}}=f^k
 \lb{51}
 \\
 {\bf f}=\frac{m}{\sqrt{1-v^2}}(-{\bf grad}\ln\xi+\xi{\bf v\times\BBB})
 \lb{52}
 \ea
 Here $\la^k_{lm}$ are the three dimensional affine connections of the gamma-space.
  As they have merely taken the equations of a space-time geodesic and shifted some terms (with a minus sign) to the other side of the equation their mathematics is unexceptionable. It is certainly true that the terms now on the right are gravitational forces and they agree with those we have derived in our special cases above, but is it clear that ALL the gravitational forces on a free particle have been taken to the right-hand side? As only gravity is acting to bend the space-time, only gravity is affecting the test particle's motion so any forces must  be gravitational. We can see that something may be amiss by considering the unphysical case with ${\bf\AAA}=0,~~\xi=1$. In such a metric the geodesics of space-time are the geodesics of space pursued at constant speed. However such spatial geodesics that come from an asymptotically flat space-time at infinity and later emerge back into it, do not do so with their momenta unchanged in direction. In dynamics  a change of momentum implies a force, however in such a metric the Landau and Lifshitz forces are zero. We deduce that even in stationary space-times there must be extra gravitational forces not included in their analysis. A specific example that demonstrates the problem is the somewhat artificial semi-Schwarzschild metric. $ds^2=dt^2-[dr^2/(1-2m/r)+r^2(d\th^2+\sin^2\th d\vf^2)]$.This disobeys the energy conditions since it has $T_0^0=0, \\~~p^1_1=~-2m/r^3,~~p_2^2=p_3^3=m/r^3;$ According to Landau and Lifshitz this metric exerts NO gravitational force on a particle moving through it. NEVERTHELESS   
  the geodesics bend through an angle of approximately $2m/b$ where $b$ is the initial impact parameter. Thus a transverse momentum has been imparted during the passage past the origin without any applied force! It may be objected that all such metrics are artificial as there can be no static deviation from flat space without a gravitational potential, however exactly the same problem occurs in real stationary spaces. In any stationary metric consider a test particle which is forced to follow a geodesic of gamma space at constant speed; the forces we have to apply to it are just those that cancel the Landau-Lifshitz forces. Thus once again the net forces are always zero yet the particle has changed its momentum during its motion through the curved part of space. Notice the locally measured curvature of its spatial path is everywhere zero since it is a spatial geodesic. We have seen a cogent example of this in the final  $v_z^2$ term in equation ($\ref{211}$). A path at constant $R,\vf$ we think of as parallel to the axis of the cylinder, but the gravity of the cylinder makes the path curved due to the $R$ dependence of the coefficient of $dz^2$ in the metric. Landau and Lifshitz treat this $v_z^2/R$ term as an acceleration due to that curvature  - not as a gravitational force. Nevertheless it is a gravitational  acceleration that is only present  due to the gravitational bending of $\ga$-space. 
  
  The problem arises because the spatial geodesics although locally as straight as possible are nevertheless globally bent.
  To alleviate this problem a global sense of direction is needed that is defined by the space itself and continuous with directions in its asymptotic flat space. As yet there is no consensus how such directions should be defined. One definition would follow gradients of Fock's "Cartesian" harmonic coordinates,  another looks to generalisations of Killing vectors $\xi^\mu$ which obey only $D^\mu(D_\mu \xi_\nu+D_\nu\xi_\mu-g_{\mu\nu}D_\la\xi^\la)=0$ or variants of that. To be useful any such definition should simplify the mathematics and lead to an extension of the concepts outside the confines of stationary space-times.  

\end{document}